\documentclass[fleqn,9pt]{wlscirep}

\usepackage{hyperref} 
\usepackage{fancyhdr}
\usepackage{lastpage}
\usepackage{setspace}
\usepackage{graphicx}
\usepackage{tocloft}
\usepackage{lmodern}
\usepackage[draft]{fixme}
\usepackage[utf8]{inputenc}
\usepackage{csvsimple}
\usepackage{amsmath}
\usepackage{float}
\usepackage{mathptmx}
\usepackage{eqnarray}
\usepackage{caption}
\usepackage{subcaption}
\usepackage{stmaryrd}
\usepackage{balance}
\usepackage{color}
\usepackage{booktabs}
\usepackage{etoolbox}
\title{On the Inability of Markov Models to Capture Criticality in Human Mobility}

\author[1*]{Vaibhav Kulkarni}
\author[2]{Abhijit Mahalunkar}
\author[1]{Benoit Garbinato}
\author[2]{John D. Kelleher}
\affil[1]{Distributed Object Programming Laboratory, UNIL-HEC Lausanne}
\affil[2]{Applied Intelligence Research Center, Dublin Institute of Technology}

\affil[*]{corresponding author: vaibhav.kulkarni@unil.ch}

\begin{document}

\begin{abstract}

\vspace{-12px}

{\bfseries

We examine the non-Markovian nature of human mobility by exposing the inability of Markov models to capture criticality in human mobility. 
In particular, the assumed Markovian nature of mobility was used to establish a theoretical upper bound on the predictability of human mobility (expressed as a minimum error probability limit), based on temporally correlated entropy. 
Since its inception, this bound has been widely used and empirically validated using Markov chains.
We show that recurrent-neural architectures can achieve significantly higher predictability, surpassing this widely used upper bound.
In order to explain this anomaly, we shed light on several underlying assumptions in previous research works that has resulted in this bias.
By evaluating the mobility predictability on real-world datasets, we show that human mobility exhibits scale-invariant long-range correlations, bearing similarity to a power-law decay.
This is in contrast to the initial assumption that human mobility follows an exponential decay. 
This assumption of exponential decay coupled with Lempel-Ziv compression in computing Fano's inequality has led to an inaccurate estimation of the predictability upper bound. 
We show that this approach inflates the entropy, consequently lowering the upper bound on human mobility predictability.
We finally highlight that this approach tends to overlook long-range correlations in human mobility. 
This explains why recurrent-neural architectures that are designed to handle long-range structural correlations surpass the previously computed upper bound on mobility predictability. 

}

\vspace{-10px}

\end{abstract}

\flushbottom
\maketitle

\section*{Introduction}

The rapid proliferation of mobile devices equipped with internet connectivity and positioning systems has resulted in the collection of massive amounts of human-mobility data.
Real-time user locations are typically collected using the Global Positioning System (GPS), Call Detail Record logs (CDR) and Wireless-LAN (WLAN); they can be used to study and model user mobility behaviours, beneficial to a variety of applications such as traffic management, urban planning and location-based advertisements. 
One of the applications of mobility modelling consists of formulating predictive models to forecast individual trajectories, for which various methods have been proposed, including Markov chains~\cite{lin2016critical}, neural networks~\cite{kulkarni:2016:MMP:3003421.3003424}, finite automata~\cite{petzold2003global} and Naive Bayes~\cite{7842898}. 
Existing research have used several datasets differing with respect to their spatial and temporal granularity, resulting in vastly contrasting prediction accuracies ranging from over 90\% to under 40\%~\cite{cuttone2018understanding}.

In this context, the seminal paper of Song et al.~\cite{Song2010LimitsOP} laid the foundations for computing a theoretical upper bound on the maximum predictability of human mobility.
This work establishes a benchmark for quantifying the performance of different algorithms and generalises its approach across various datasets.    
It defines the \emph{maximum predictability}, noted~$\pi^{max}$, as the temporally correlated entropy of information of an individual's trajectory.
$\pi^{max}$ is estimated by first computing the entropy based on the Lempel-Ziv data compression~\cite{ziv1978compression} and then by solving the limiting case for Fano's inequality~\cite{prelov2008mutual}, an information-theoretical result used to compute lower bounds on the minimum error probability in multiple-hypotheses testing problems.
The proposed theoretical upper bound ($\pi^{max} = 93\%$) is empirically validated using a CDR dataset collected over 50,000 users by a telecommunications  operator for a duration of three months.
It should be noted however that CDRs are a rather rough approximation of  human mobility.
Song et al.~\cite{Song2010LimitsOP} also show that $\pi^{max}$ is independent of radius of gyration and movement periodicity, hence observe an insignificant level of variation across a heterogeneous population.

Several subsequent works either redefine this theoretical upper bound or perform empirical validations with different mobility datasets.
Lu et al.~\cite{lu2013approaching} re-estimate $\pi^{max}$ to be 88\% and use Markov chains to empirically verify this redefined upper bound. 
They analyse another CDR dataset consisting of 500,000 users, collected for a duration of five months and achieve an average predictability of $91\%$ with an \textit{order-1} Markov chain.
They also show that higher-order Markov chain models do not improve prediction accuracy. 
Their interpretation behind surpassing their own estimated theoretical bound is that trajectories exceeding this bound are non-stationary, whereas the accuracy of stationary trajectories prevails within the bound. 
A trajectory is considered to be stationary when people tend to remain still during short time-spans.
This conclusion directly contradicts findings of Song et al.~\cite{Song2010LimitsOP}, because non-stationary trajectories should by definition have a higher entropy.
Additionally, Cuttone et al.~\cite{cuttone2018understanding} show that the stationary nature of trajectories plays a significant role in the higher accuracies resulting from Markov models~\cite{cuttone2018understanding} as they often predict the user will remain in the previous location, i.e., self-transitions.
Lin et al.~\cite{lin2016critical} also show that $\pi^{max}$ is independent of the data sampling rate which was later questioned by Smith et al.~\cite{smith2014refined} ($\pi^{max} = 81 \pm 4\%$ ) and Cuttone et al.~\cite{cuttone2018understanding} ($\pi^{max} = 65\%$).
Smith et al.~\cite{smith2014refined} and Cuttone et al.~\cite{cuttone2018understanding} use mobility datasets~\cite{zheng2010geolife, stopczynski2014measuring} containing GPS trajectories and empirically show that predictability has a direct correlation with the temporal resolution and an inverse correlation with the spatial resolution.

The CDR datasets used in these works are known to have inherent gaps due to the short bursts of calls masking the user's true entropy.
Human mobility varies under time translations, therefore the entropy not only depends on the duration of past observations but also on number of visited locations, these factors tend to be hidden in such datasets~\cite{barabasi2005origin, bialek1999predictive}. 
Additional inconsistencies become evident due to the fact that the authors in~\cite{Song2010LimitsOP, lu2013approaching} group the user locations into one hour bins when constructing the historical trajectory of a user.
Further inspection suggests that these models can foresee future locations at $\pi^{max}$, only when an individual is present in one of the top \textit{n} bins~\cite{cuttone2018understanding}.
The first two works~\cite{Song2010LimitsOP, lu2013approaching} thus consider the last location of each day, consequently predicting only the user's home place.
Under such a scenario, Ikanovic et al.~\cite{ikanovic2017alternative} and Cuttone et al.~\cite{cuttone2018understanding} show that the predictability of the true next location is significantly lower ($\pi^{max}=71.1 \pm 4.7\%$) than the predictability of the location in the subsequent bin.
They further show that an individual's mobility entropy is directly proportional to the number of visited locations. 
The authors also point out that the generating function behind the stochastic mobility behaviour is often unknown.
Therefore the bounds cannot be estimated theoretically and require empirical derivation. 
Cuttone et al.~\cite{cuttone2018understanding} achieve an even lower bound on $\pi^{max}$ of 65\% on the same datasets with the same methods as Ikanovic et al.~\cite{ikanovic2017alternative}.

Zhao et al.~\cite{zhao2015non} argue that lower entropy and higher predictability in mobility behaviour originates from its non-Markovian character.
To highlight the non-Markovian character of human mobility, they analyse the rank distribution of the visited locations and the associated dwelling times.
Using a CDR dataset, they show the presence of the scaling law (Zipf's) behaviour in one-point statistics~\cite{newman2005power}. 
However, human mobility is better described by two-point statistics as it involves implicit behavioural dynamics while traveling from a given location to another with a specific intent~\cite{lin2016critical}.

Based on the research literature discussed above, we observe a discrepancy regarding the maximum predictability bound, $\pi^{max}$ and disagreements on the impact of entropy, the number of uniquely visited locations and the spatiotemporal resolution of the trajectory on $\pi^{max}$.   
In order to gain a deeper understanding about this discrepancy, we construct next-place prediction models by using seven different approaches.
We compute the empirical maximum accuracy and compare it with the theoretically derived $\pi^{max}$, considering three large-scale real-world datasets containing GPS trajectories. 
We find that recurrent-neural architectures~\cite{schmidhuber2015deep} significantly surpass $\pi^{max}$ on datasets compiled for long timespans.

The current approaches~\cite{Song2010LimitsOP, lin2016critical} estimate $\pi^{max}$ by first computing the true entropy of user mobility, denoted by $S^{real}$ using Lempel-ziv compression~\cite{ziv1978compression}, followed by the computation of the minimum error probability by leveraging Fano's inequality~\cite{ho2008conditional, prelov2008mutual}.
Fano's inequality is based on the assumption that the system is governed by a Markovian process and computes the conditional entropy of a random variable $X$, relative to the correlated variable $Y$, to the probability of incorrectly estimating $X$ from $Y$, thus yielding the minimum error probability, noted $p_e$.
In practice, $p_e$ is computed by segmenting the entire trajectory into sub-strings, where the length of the shortest substring beginning at index $i$ does not appear previously~\cite{lu2013approaching, Song2010LimitsOP, smith2014refined, ikanovic2017alternative}.
The total predictability $\pi^{max}$ is thus the complement of the average of error probabilities on individual substrings.

We show that when a trajectory is split into substrings, the entropy associated with the individual substring increases, thus increasing $S^{real}$.
This occurs as Fano's inequality, rooted in information theory~\cite{arimoto1971information}, is intended for a data source with known probability distribution~\cite{gerchinovitz2017fano}.
Human mobility prediction however, is based on the discretization of the trajectories, where the probability distribution is not known {\it{a priori}}.
Furthermore, the estimation of entropy by using Lempel-Ziv coding~\cite{ziv1978compression} was originally constructed to provide a complexity measure for finite sequences, i.e., input sequence displaying exponential decay in long-range correlations (memoryless structure). 
Such sequences, when further split, changes the true distribution of the data and increase the associated entropy; and the derived $\pi^{max}$ thus acts as a limit on the Markov model.
We argue that this is due to ignoring the presence of long-range structural correlations that are present in human trajectories.

The current $\pi^{max}$ computation is thus based on the widely used assumption that human mobility is Markovian (memoryless), i.e., the movements are independently distributed, as testified by the numerous research works relying on Markov models to characterize mobility~\cite{lu2013approaching, song2006evaluating, krumme2013predictability, bapierre2011variable, kulkarni:2016:MMP:3003421.3003424}.
In opposition to this widely used assumption, we show the presence of non-Markovian character in human mobility dynamics by empirically showing that the drop in the mutual information~\cite{prelov2008mutual, cflsp} follows a power law function of the distance between any two points in a trajectory.
We also show the state-dependent nature of mobility by illustrating the presence of scaling laws in the distribution of location dwell times and the rank associated with location visits. 
Using real-datasets, we show that these underlying mechanisms that govern human mobility are the same across disparate mobility behaviours.

\section*{Background}

\subsection*{Complexity and Criticality}

Self-organized complexity is a phenomenon which combines self-organization and criticality to describe complexity~\cite{bak1987self}.
It is a property of dynamic systems to regulate their microscopic behavior to be spatial/temporal scale independent~\cite{turcotte1999self}.
This resembles their behavior at a critical point of phase transition.  
However, unlike conventional phase transitions, self-organized complexity governed systems do not depend on any external tuning of control parameters, i.e., the evolution of dynamic complex systems is self organized into the critical behavior~\cite{bak1987self,turcotte1999self}.\\

The critical point of a statistical system is defined as the point in space, parameterized by intensive quantities (for example, temperature and pressure), at which there exist no boundaries between the phase transitions.
Simply put, it is the end of the co-existence curves of two phases~\cite{rice1955shape}.
On the exterior of this critical point, the different phases are cramped due to a correlation in the phase (or other quantifiable properties) of adjacent elements. 
Such a correlation prevails at come distinctive length scale (the correlation length).
This can be also visualized as the density of the boundary between the phases. 
When a dynamic system arrives at the critical point, this correlation length becomes infinite, resulting in fuzzing the boundaries. 
Such an infinite correlation length does not contain any information regarding a finite scale length.
As one cannot use an infinite length as a unit of measurement, the theory describing the system behavior at the critical point is scale-invariant.
Therefore, it is clear that critical behavior implies scale invariance, and therefore when critical behavior is involved, the effects at distances much longer than microscopic lengths are imperative to study.\\

In general, a dynamic system self organizes into a complex state but with a fairly general structure. 
The complexity occurs due to an absence of a single well defined characteristic event size. 
However, despite the complexity, the system exhibits some statistical properties governed by power laws~\cite{newman2005power}.
For instance, the number of events, $E$ as a function of its size $s$ (wherein a major event is less likely to occur as compared to a small one) can be described as in Equation~\ref{eq:eq1}.

\begin{equation}
\label{eq:eq1}
E(s) = A\times s^{-\alpha}
\end{equation}

Where $A$ is some constant and $\alpha$ describes the statistical features of a self-organized critical (SOC) state.
In general SOC has been observed in slowly-driven and non-equilibrium systems which posses extended degrees of freedom and high level of non-linearity~\cite{turcotte1999self,jensen1998self}.
In this paper, we show that human mobility patterns follow the above characteristics.\\

\subsection*{Criticality in Practice}

We give some examples to connect such a physical phenomenon to instances in the practical world. 
There has not been a plethora of work to characterize practical systems as SOC.
Amongst the limited work, we look at how Carreras et al.~\cite{carreras2004evidence} analyze a 15-year long time series of electric power transmission system blackouts.
The authors present a possible quantitative explanation of the complex dynamics in a power systems which contains properties which leads to dynamic equilibrium with some properties of SOC.
Another work by Bartolozzi et al.~\cite{bartolozzi2005self} consider the stock markets which are complex self-interacting systems, characterized by intermittent behaviors. 
Here periods of high activity alternate with periods of relative calm. 
On the basis on empirical evaluations the authors conclude that stock markets depict a near-SOC state.

In physical systems, criticality occurs when there exists correlation length-scale or time-scale that diverges to infinity. 
Therefore, the relationships in the sequence are not local in time, but are arbitrarily non-local.

\subsection*{Exponential Decay and Power-law}

In SOC dynamic systems, the probability of occurrence of major disruptive events decreases as a power function of the event size~\cite{bak1987self}. 
This is in contrast to many conventional systems in which this probability decays exponentially with event size.
The difficulty arises as the correlation length is infinite.
However, it is well known and established that in a self-organized critical system (SOC), the long range interactions occur due to power-law decay of correlation.\\

The power law is detected and characterized, by combining maximum-likelihood fitting methods with goodness-of-fit tests based on Kolmogorov-Smirnov statistic and likelihood ratios~\cite{clauset2009power}. 

At this point, we summarize the difference between exponential decay and power law.
Although visually both might look similar as both are positive and go asymptotically to 0, the difference is pointed out in Equation~\ref{eq:eq2} and Equation~\ref{eq:eq3}.

\begin{equation}
	\label{eq:eq2}
	Power \: Law: y = x^{k}
\end{equation}

\begin{equation}
	\label{eq:eq3}
	Exponential \: Decay: y = k^{x}
\end{equation}

Where $k$ is some constant.
Thus, it is evident that exponential decay goes to zero much faster than power law.
Furthermore, exponential probability distribution has an inherent property of memoryless-ness, whereas power law follows a polynomial relationship exhibiting the property of scale invariance.

\subsection*{Entropy \& Mutual Information}

Formally, the entropy of a discrete random variable $X$ with probability mass function (pmf) $p_{X}(x)$ is:

\begin{equation}
	\label{eq:eq4}
	H(X) = -\sum_{x} p(x)\log p(x) = -E[\log(p(x))]
\end{equation}

In a nutshell, the entropy measures the expected uncertainty in $X$.
Therefore, $H(X)$ is equal to the amount of information learnt on an average from one instance of the random variable $X$ and $p(x)$, the probability distribution.
Furthermore, the entropy does not depend on the value that the random variable takes, but only on the probability distribution $p(x)$~\cite{shannon2001mathematical}.

In case two random variables, $X,Y$ jointly distributed according to the $pmf$ $p(x,y)$, the joint entropy is given by:

\begin{equation}
	\label{eq:eq5}
	H(X,Y) = -\sum_{x,y}p(x,y)\log p(x,y)
\end{equation}

Furthermore, the conditional entropy of $X$ given $Y$ can be given by:

\begin{equation}
	\label{eq:eq6}
	H(X|Y) = -\sum_{x,y}\log p(x|y) = -E[\log(p(x|y))]
\end{equation}

Simply put, the conditional entropy is a measure of how much uncertainty remains about a random variable $X$ given $Y$. 

The {\bfseries{Mutual Information}} $I$, between two discrete random variables $X,Y$ jointly distributed according to $pmf$ $p(x,y)$ is given by:

\begin{equation}
\begin{aligned}
\label{eq:eq8}
I(X;Y) &= \sum_{x,y}p(x,y)\log\frac{p(x,y)}{p(x).p(y)} \\
& = H(X) - H(X|Y) \\
& = H(Y) - H(Y|X) \\
& = H(X) + H(Y) - H(X,Y)
\end{aligned}
\end{equation}

Mutual information can be analogously defined for continuous variables and conditional mutual information.
Simply, mutual information can be explained at follows.
Consider two people Alice and Bob, living in Switzerland. 
Alice goes out in a t-shirt if the temperature is more than 20 degrees celsius.
Bob always wears a t-shirt irrespective of the temperature.
Now, notice that Alice's actions give information about the temperature in Switzerland.
However, Bob's actions give no information whatsoever. 
This is due to Alice's actions are random and correlated with the temperature, whereas Bob's actions are deterministic. 
Mutual information $I$ mathematically quantifies this notion.

\subsection*{Markov Processes}

{\noindent{{\bfseries{Memoryless Markov Models.}}}} A Markov process is defined by a matrix $M$ containing conditional probabilities $M_{ab}$ such that~$M_{ab} = P(X_{t+1} = a | X_t = b)$.
The Markov matrices (or stochastic matrices) are bounded by $M_{ab} \geq 0$ and $\sum_{a}M_{ab} = 1$.
Therefore the complete dynamic of the model can be specified as provided in equation below:

\begin{equation}
	p_{t+1} = M.p_{t}
\end{equation}

where $p_t$ is a vector with components $P(X_t = a)$ that specifies the probability distribution at time $t$.
If $\lambda_i$ denotes the eigenvalues of $M$, sorted in decreasing magnitude such that: $|\lambda_1| \geq |\lambda_2| \geq |\lambda_3|$.
The Markov matrices have $|\lambda_i| \leq 1$, with corresponding eigenvector resulting in a stationary probability distribution.

Furthermore, two additional conditions are imposed on Markov matrices: i) $M$ is irreducible, i.e. every state is accessible from every other state, and ii) $M$ is aperiodic, to avoid processes that will never converge (for example $a \rightarrow b \rightarrow a \rightarrow b$). Thus it has been shown that $M$ being irreducible and aperiodic implies that the individual eigenvalues are less than one and the stationary probability distribution is unique~\cite{rigoperron}.\\

{\noindent{{\bfseries{Hidden Markov Models (HMM).}}}} In an HMM, in addition to the observed sequences $X_1, X_2,...,X_n$, there are hidden latent variables $Y_1, Y_2...,Y_n$ that form a Markov chain as well.
These internal hidden dynamics are never observed, but at each time-step, an output is produced such that $Y_i \rightarrow X_i$ form the observed sequence.\\

{\noindent{{\bfseries{Markov Model with Memory.}}}} Markov processes are able to model and represent higher-order dependencies between successive observations of a state variable.
One way to increase the process memory is by using high-order Markov models.
In such a model, the conditional probability distribution of future states in the process not only depends on the current state but also on the past states.
A Markov chain with memory $m$ is a process satisfying,

\begin{equation}
	P(X_{t+1 = s_{t+1}|X_t = s_t,...,X_1=s_1}) = P(X_{t+1} = s_{t+1}|X_t=s_t,...,X_{t-m+1} = s_{t-m+1})
\end{equation}

for all $t \geq m =$. By defining

\begin{equation}
	Y_t = (X_t, X_{t-1},...,X_{t-m+1})
\end{equation}

and by taking the ordered $m-tuples$ of $X$ values as its product space so that the chain ${Y_t}$ with suitable starting values satisfies the Markov property.

\section*{Results}

In this section, we first present the values of $S^{real}$ and $\pi^{max}$ computed using the approach mentioned in the works of Song et al.~\cite{Song2010LimitsOP} and Lu et al.~\cite{lu2013approaching}.
We then discuss the accuracy results estimated using seven algorithms and compare them with respect to the theoretical upper bound.
To investigate the presence of memory in human mobility, we conduct several experiments and use the results to illustrate the reason for surpassing the upper bound. 
Finally, we investigate the existing approach and discuss their failure to compute the true entropy ($S^{real}$) of mobility trajectories.\\

\noindent {\bfseries{Mobility datasets.}} We conduct the experiments by using three mobility datasets whose specifications, along with the estimated values of $S^{real}$ and $\pi^{max}$, are shown in Table~\ref{tbl:tbl1}. 
The PrivaMov dataset~\cite{mokhtar2017priva} was collected through GPS, WiFi and GSM in the city of Lyon (France) and includes university students, staff and their family  members.
The Nokia mobile dataset~\cite{laurila2012mobile} (NMDC) was collected in the Lake Geneva region of Switzerland and consists young individuals' trajectories, collected through GPS, WLAN, GSM and Bluetooth.
The GeoLife dataset~\cite{zheng2010geolife} was collected in Beijing (China) and contains trajectories recorded through GPS loggers and GPS-phones.
The computation of $S^{real}$ and $\pi^{max}$ at the aggregate level for the dataset is based on our observation of independence of predictability on travel distance (radius of gyration $r_g$) in human mobility, which is consistent with previous studies~\cite{lu2013approaching, Song2010LimitsOP, yan2013diversity}.

\begin{table}[h!]
\resizebox{\textwidth}{!}{%
\begin{tabular}{|c|c|c|c|c|c|c|c|}
\hline
Datasets & \textbf{\begin{tabular}[c]{@{}c@{}}Num.\\ users\end{tabular}} & \textbf{\begin{tabular}[c]{@{}c@{}}Duration\\ (months)\end{tabular}} & \textbf{\begin{tabular}[c]{@{}c@{}}Avg. trajectory\\ length ($n$)\end{tabular}} & \textbf{\begin{tabular}[c]{@{}c@{}}Distinct locations\\($N$)\end{tabular}} & \textbf{\begin{tabular}[c]{@{}c@{}}Avg. spatio-temporal\\ granularity\end{tabular}} & {\bfseries{$S^{real}$}} & {\bfseries{$\pi^{max}$}} \\ \hline
\textbf{PrivaMov} & 100 & 15 & 1560000 & 2651 & \begin{tabular}[c]{@{}c@{}}246 meters\\ 24 seconds\end{tabular} & 6.63 & 0.5049 \\ \hline
\textbf{NMDC} & 185 & 24 & 685510 & 2087 & \begin{tabular}[c]{@{}c@{}}1874 meters\\ 1304 seconds\end{tabular} & 5.08 & 0.6522 \\ \hline
\textbf{GeoLife} & 182 & 36 & 8227800 & 3892 & \begin{tabular}[c]{@{}c@{}}7.5 meters\\ 5 seconds\end{tabular} & 7.77 & 0.4319 \\ \hline
\end{tabular}%
}
\caption{Dataset specifications along with their respective $S^{real}$ and $\pi^{max}$.}
\label{tbl:tbl1}
\end{table}

\begin{figure}[h!]
    \centering
      \begin{subfigure}[b]{0.32\textwidth}
        \includegraphics[scale=0.105]{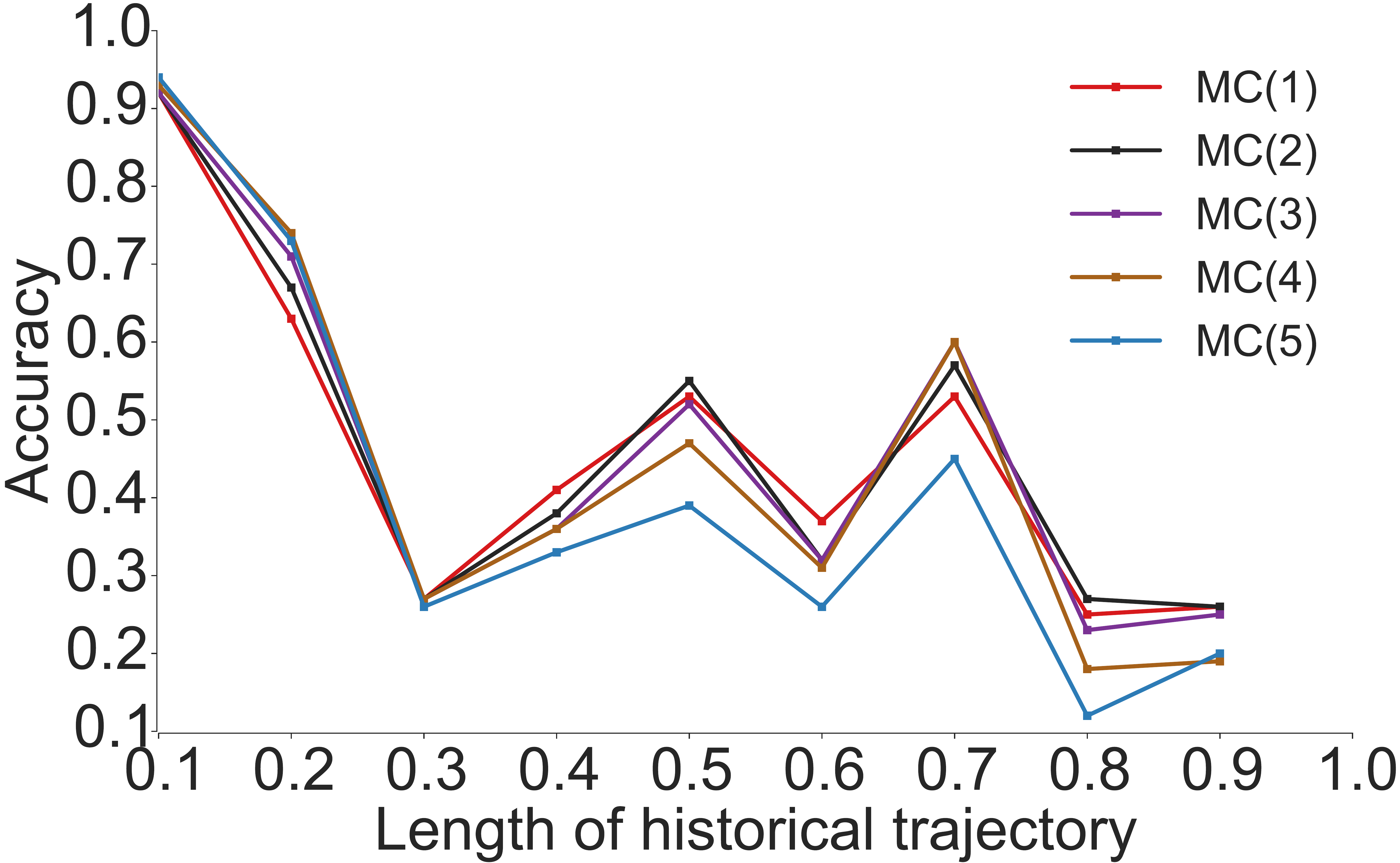}
        \caption{PrivaMov dataset}
        \label{fig:markov_privamov}
    \end{subfigure}
    \begin{subfigure}[b]{0.32\textwidth}
        \includegraphics[scale=0.105]{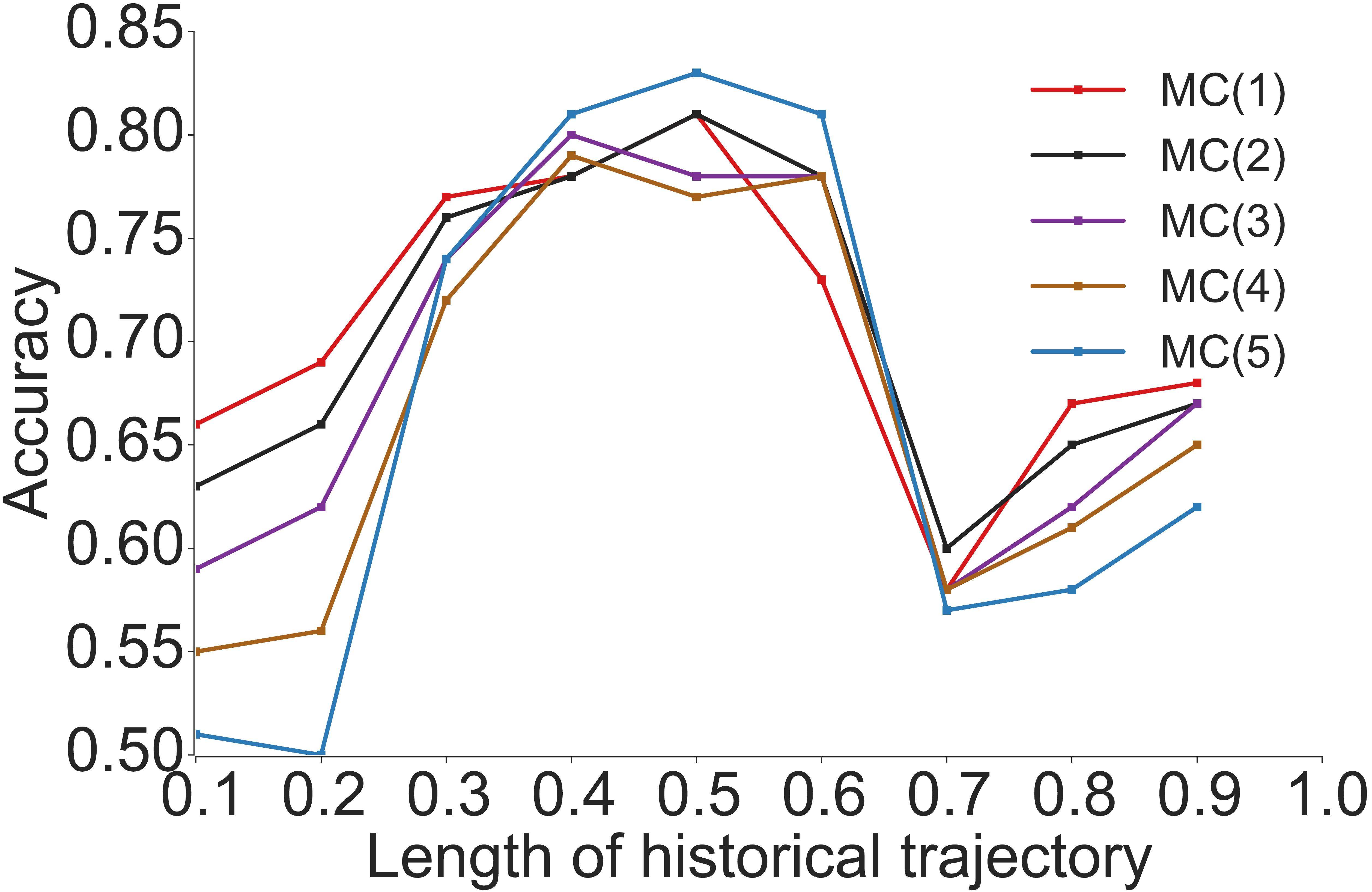}
        \caption{NMDC dataset}
        \label{fig:markov_nmdc}
    \end{subfigure}
        \begin{subfigure}[b]{0.32\textwidth}
        \includegraphics[scale=0.105]{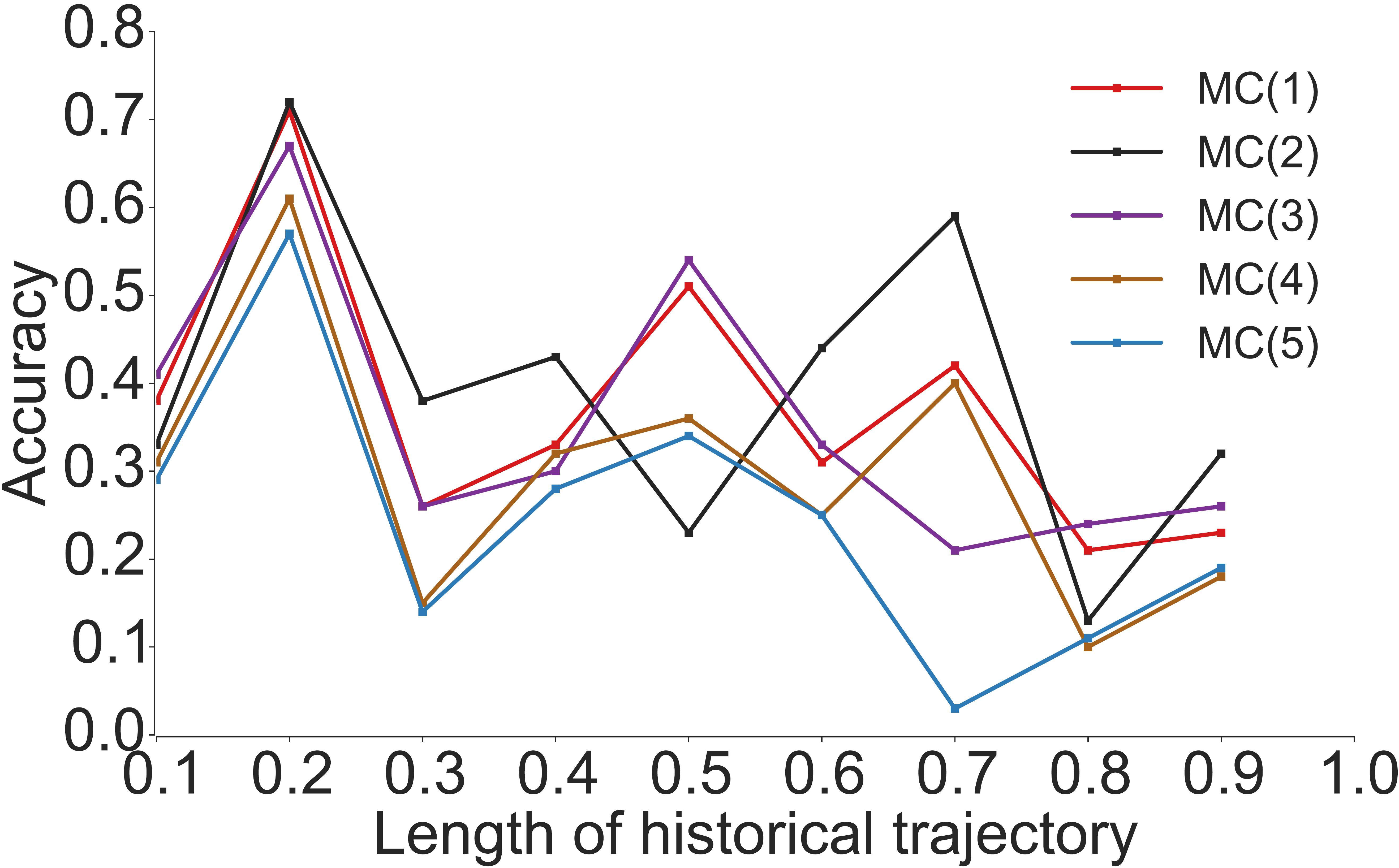}
        \caption{GeoLife dataset}
        \label{fig:markov_geolife}
    \end{subfigure}
    \caption{Prediction accuracy for Markov models (order 1-5).}
    \label{fig:markov_all_datasets}
\end{figure}

\begin{figure}[h!]
    \centering
      \begin{subfigure}[b]{0.32\textwidth}
        \includegraphics[scale=0.105]{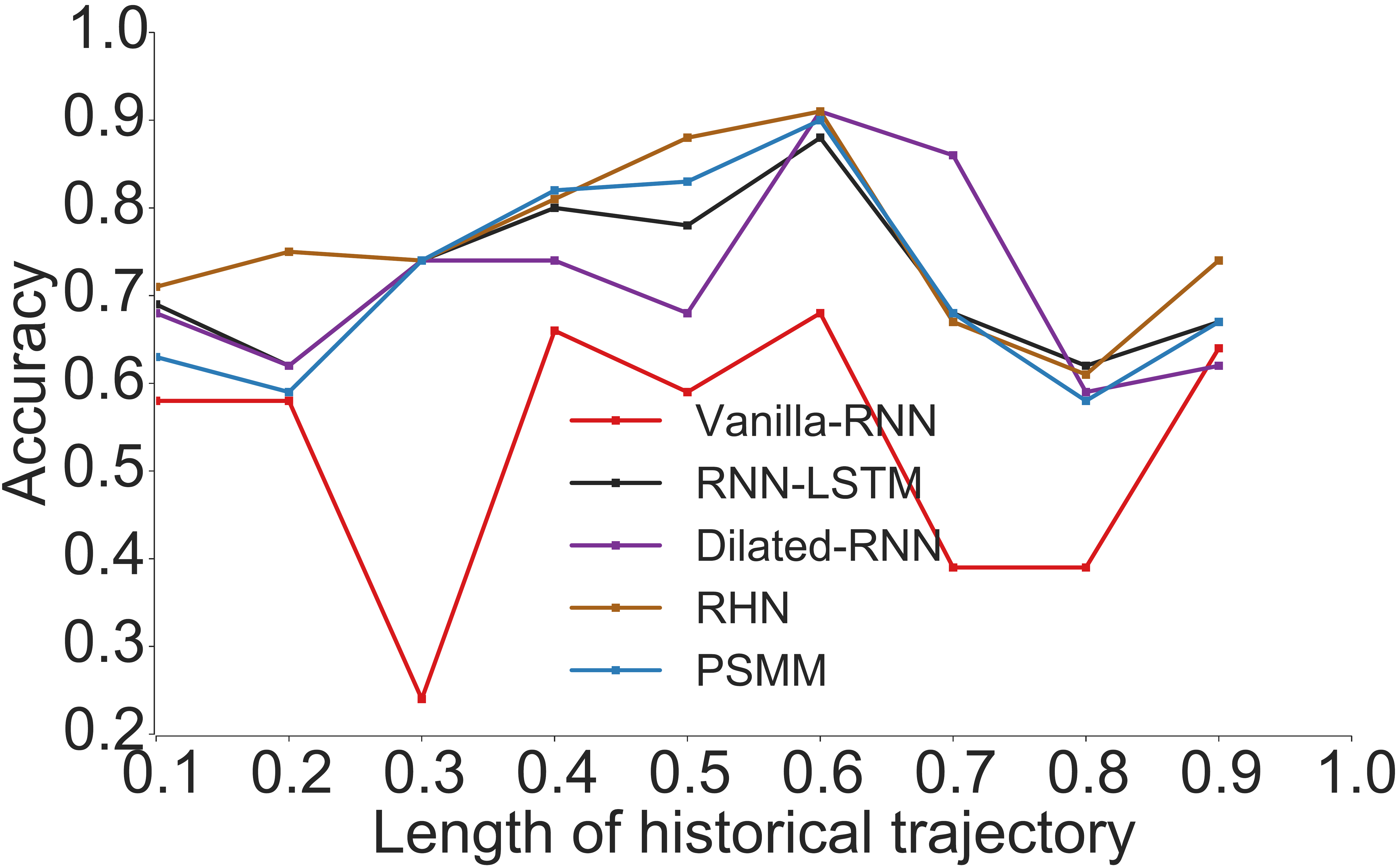}
        \caption{PrivaMov dataset}
        \label{fig:a1}
    \end{subfigure}
    \begin{subfigure}[b]{0.32\textwidth}
        \includegraphics[scale=0.105]{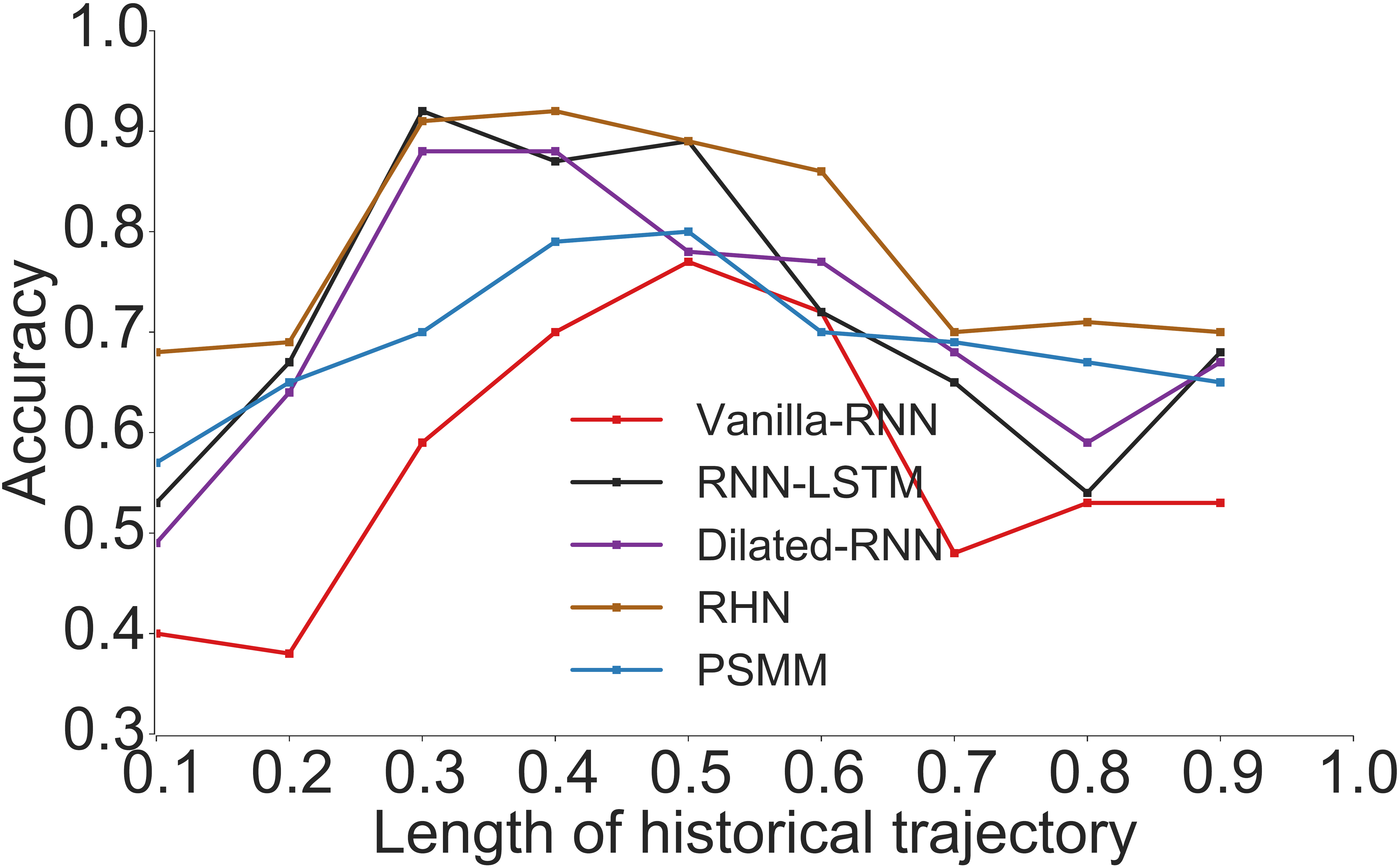}
        \caption{NMDC dataset}
        \label{fig:a2}
    \end{subfigure}
        \begin{subfigure}[b]{0.32\textwidth}
        \includegraphics[scale=0.105]{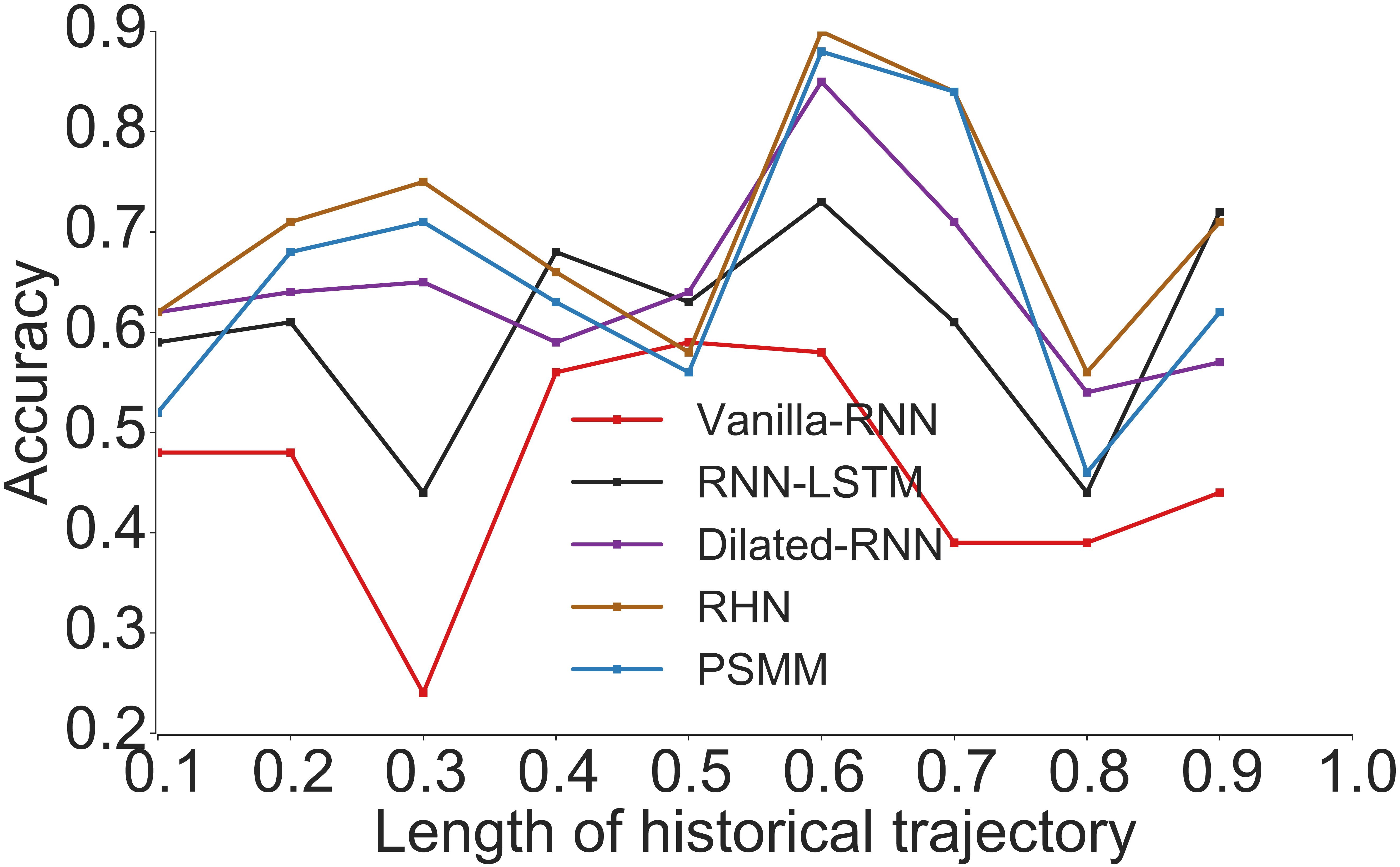}
        \caption{GeoLife dataset}
        \label{fig:a3}
    \end{subfigure}
    \caption{Prediction accuracy for recurrent-neural architectures.}
    \label{fig:memory_nets_all_datasets}
\end{figure}

\noindent {\bfseries{Prediction Algorithms.}} We estimate the empirical predictability using seven different approaches: (1) Markov chains~\cite{gambs2012next} (order 1-5), (2) Hidden Markov model~\cite{si2010mobility}(HMM), (3) Vanilla Recurrent Neural Network~\cite{grossberg2013recurrent} (Vanilla-RNN), (4) Recurrent Neural Network with Long Short-Term Memory~\cite{Hochreiter1997LongSM} (RNN-LSTM), (5) Dilated Recurrent Neural Network~\cite{Chang2017DilatedRN} (Dilated-RNN), (6) Recurrent Highway Network~\cite{Zilly2017RecurrentHN} (RHN), and (7) Pointer Sentinel Mixture Model~\cite{Merity2016PointerSM} (PSMM).
We find that higher order Markov chains (typically $>3$) do not contribute to increase prediction accuracy, as also observed by Lu et al.~\cite{lu2013approaching}.
The prediction accuracy for Markov chain models and the recurrent-neural architectures for all datasets is shown in Figure~\ref{fig:markov_all_datasets} and Figure~\ref{fig:memory_nets_all_datasets}, respectively.

\begin{figure}[h!]
    \centering
      \begin{subfigure}[b]{0.32\textwidth}
        \includegraphics[scale=0.103]{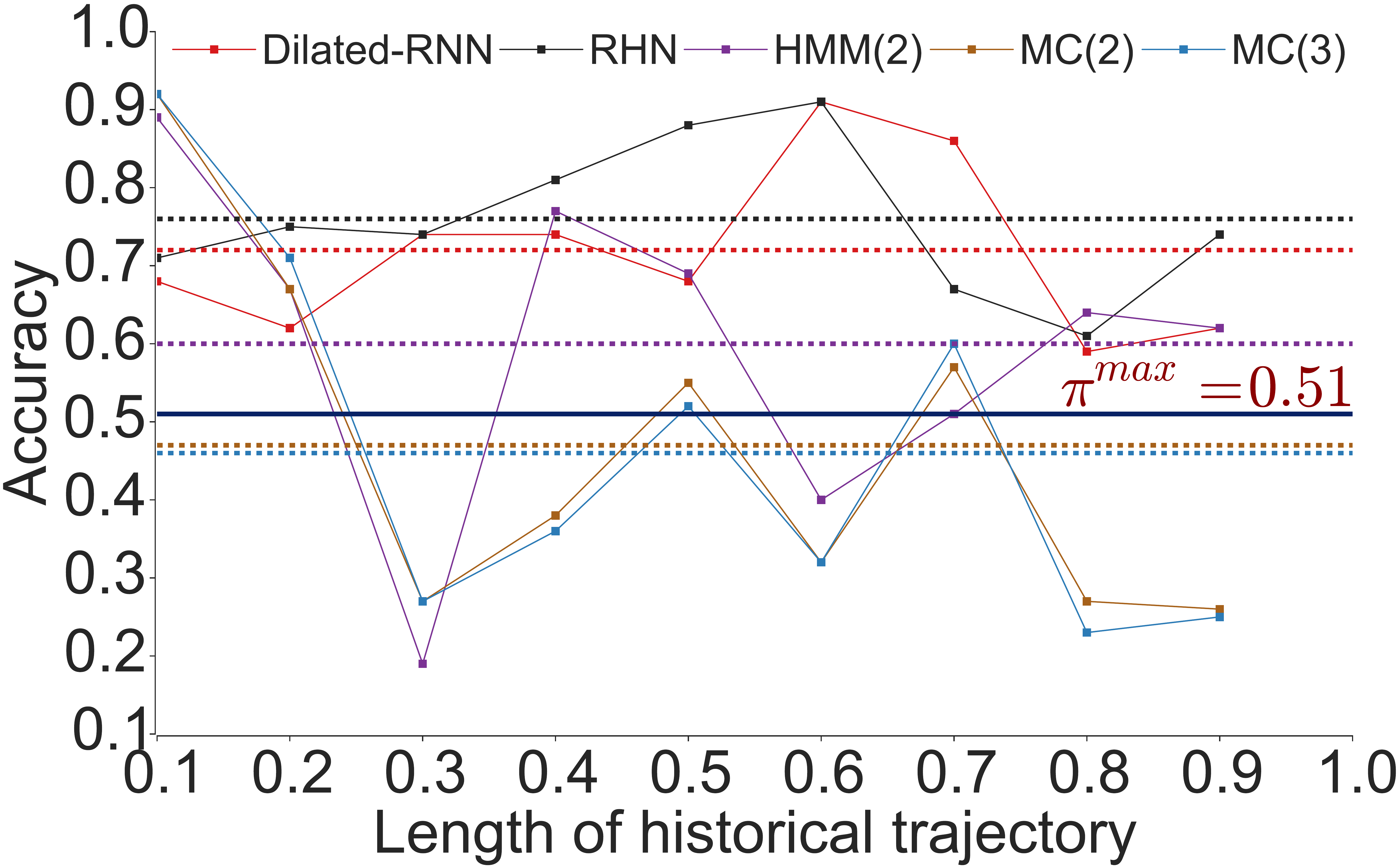}
        \caption{PrivaMov dataset}
        \label{fig:a1}
    \end{subfigure}
    \begin{subfigure}[b]{0.32\textwidth}
        \includegraphics[scale=0.103]{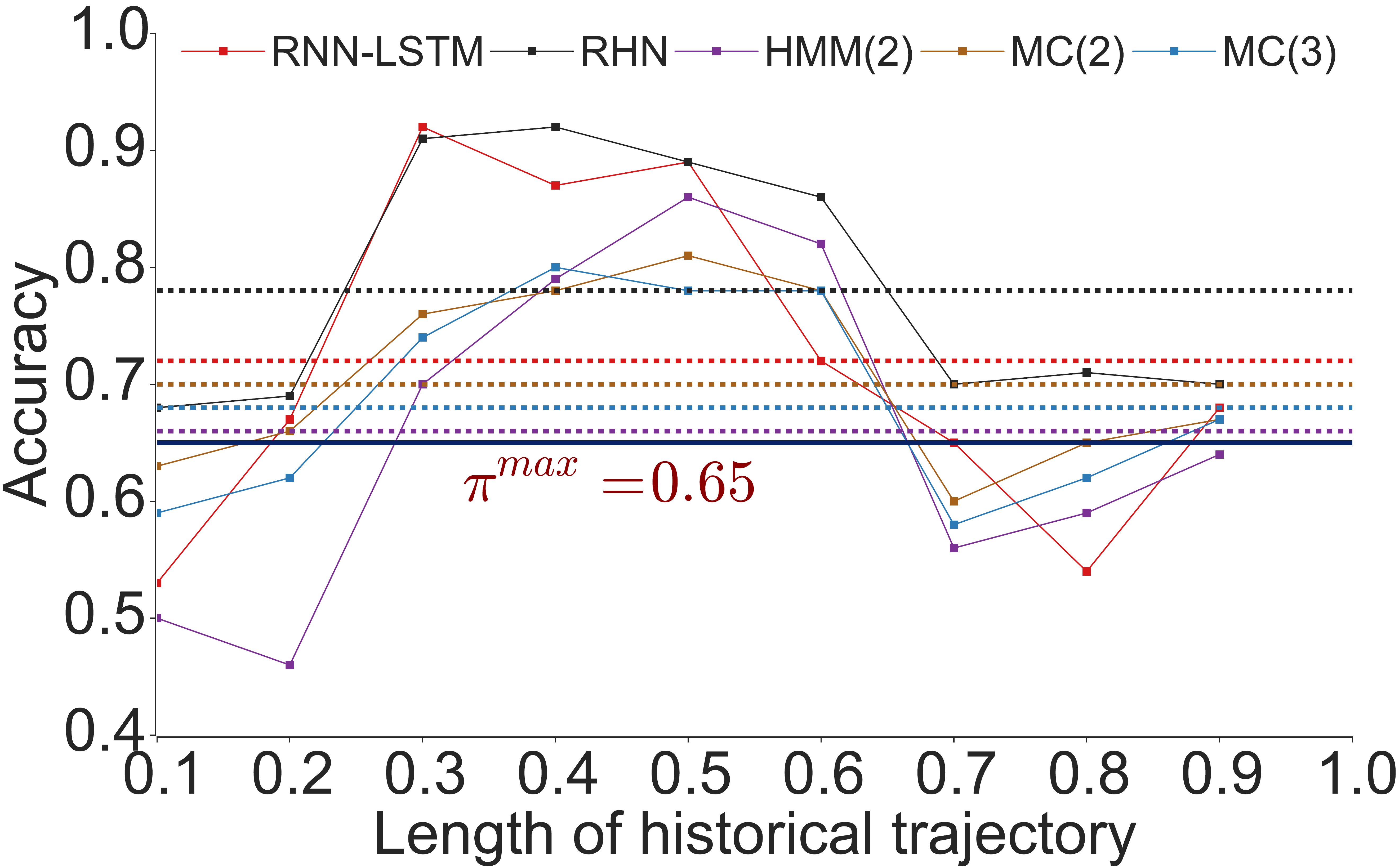}
        \caption{NMDC dataset}
        \label{fig:a2}
    \end{subfigure}
        \begin{subfigure}[b]{0.32\textwidth}
        \includegraphics[scale=0.103]{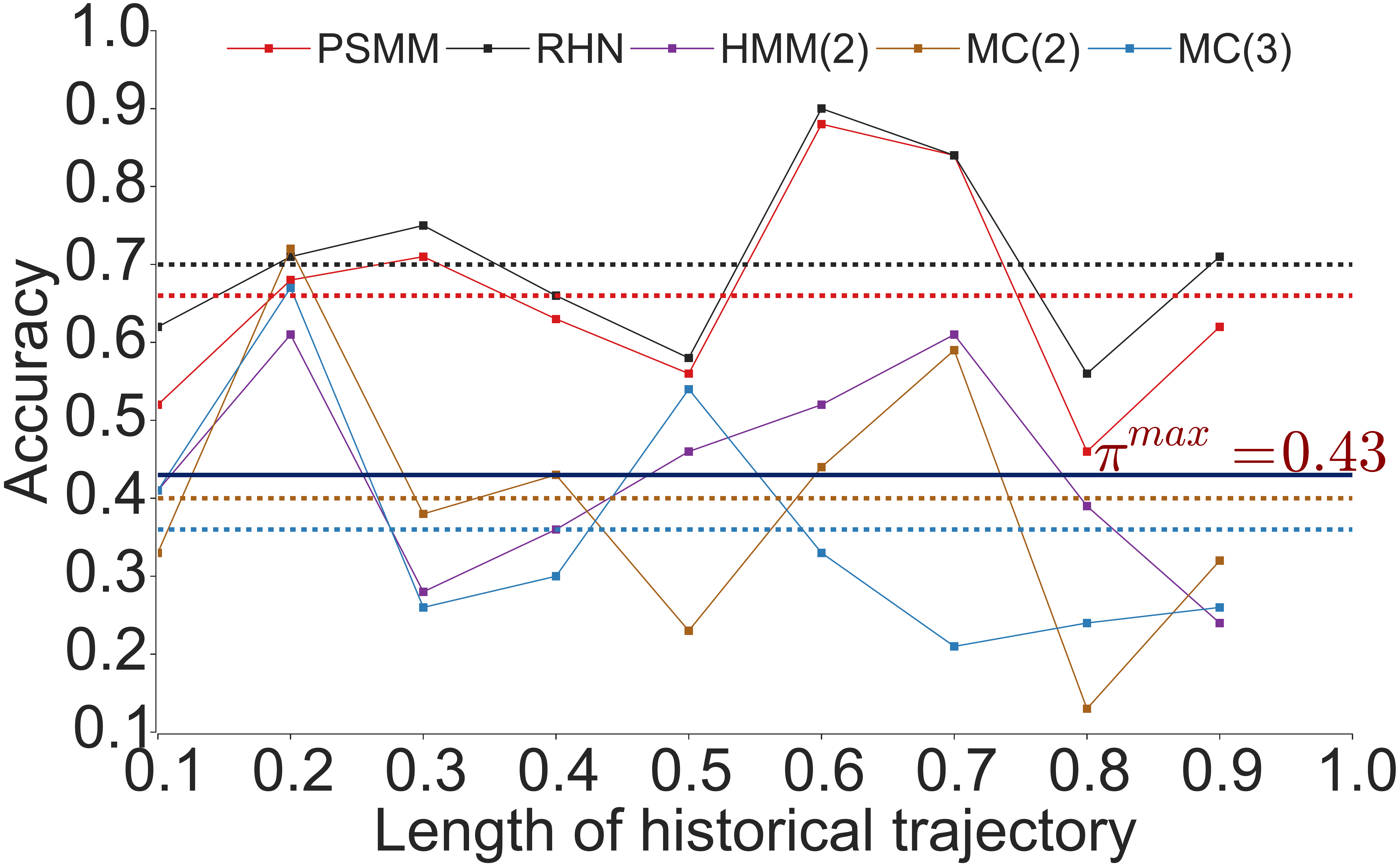}
        \caption{GeoLife dataset}
        \label{fig:a3}
    \end{subfigure}
    \caption{Comparison of $\pi^{max}$ with the maximum predictability achieved using models from each category. The dotted lines indicate the predictability by each approach (indicated with the same colour).}
    \label{fig:comparison_pi_max}
\end{figure}

We observe that the accuracy of Markov models lie in the vicinity of the theoretical $\pi^{max}$. 
It is also clearly evident that recurrent-neural architectures significantly outperform Markov models with respect to their average accuracies.
Recurrent-neural architectures are a class of artificial neural networks, which use their hidden memory representation to process input sequences. 
The variants of these architectures differ in their capacity to manipulate this memory and propagate gradients along the network.
For instance, RHN's are built to account for short and long-range correlations present in a sequence, which explains their superior performance as compared to the other architectures. 
Whereas, PSMM's weigh long-range dependencies much higher than short-distance correlations in the sequence.
In Table~\ref{tbl:tbl2}, we show the maximum predictability achieved by using the best performing models from each algorithm, and in Figure~\ref{fig:comparison_pi_max} we compare their performance with the theoretical upper bound.\newline

\begin{table}[h!]
\centering
\begin{tabular}{|c|c|c|c|c|c|}
\hline
Datasets          & \textbf{$\pi_{MC(2)}^{max}$} & \textbf{$\pi_{MC(3)}^{max}$} & \textbf{$\pi_{HMM}^{max}$} & \textbf{$\pi_{RHN}^{max}$} & \textbf{$\pi_{RNN}^{max}$} \\ \hline
\textbf{PrivaMov} & 0.47                  & 0.46                  & 0.60                   & 0.76                & 0.72 (Dilated-RNN)  \\ \hline
\textbf{NMDC}     & 0.70                  & 0.68                  & 0.66                   & 0.78                & 0.72 (RNN-LSTM)     \\ \hline
\textbf{GeoLife}  & 0.40                  & 0.36                  & 0.43                   & 0.70                & 0.66 (PSMM)         \\ \hline
\end{tabular}
\caption{Prediction accuracy achieved using the best performing models for each dataset.}
\label{tbl:tbl2}
\end{table}

The prediction accuracies of recurrent-neural architectures also surpass the theoretical upper bound for the respective dataset.  
This anomaly in computing $\pi^{max}$ is puzzling, even more so considering the diversity of the datasets with respect to their collective time spans, visited number of locations, demographics and spatiotemporal granularity.
This lets us question the assumption that human mobility follows a Markov process. 
Here, we conduct extensive analysis to empirically prove its non-Markovian nature.\\

\begin{figure}[t!]
\centering
\includegraphics[scale=0.15]{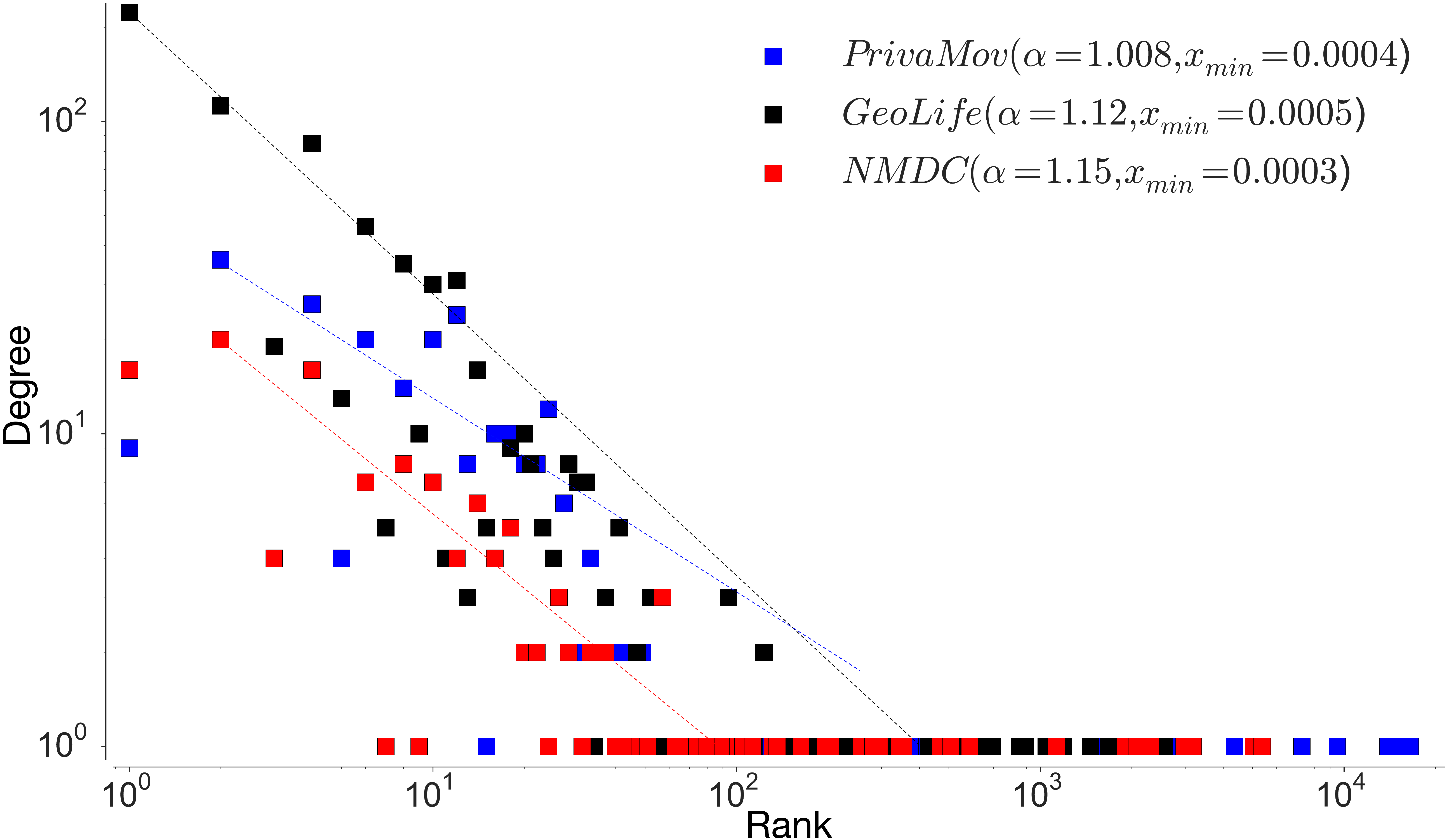}
\caption{Rank distribution of location visits at the collective level for aggregated dataset. The data is binned into exponentially wider bins and normalised by the bin width. The straight line represents the fitting through least squares regression ($\alpha$ and $x_{min}$, computed through maximum likelihood estimation).}
\label{fig:rank_distribution}
\end{figure}

\noindent {\bfseries{Location Rank Distribution.}} In order to gain insight into the datasets, we first analyse the rank distribution of the locations, according to the visit frequency at individual and aggregated levels.
An individual visits different locations depending on a perceived priority attached to the location~\cite{barabasi2005origin}; this results in a heterogenous location frequency distribution~\cite{zhao2015non}.
To study the location-rank distribution, we follow the approach stated in Zhao et al.~\cite{zhao2015non} in order to rank locations according to their collective magnitude at the aggregate level. 
Figure~\ref{fig:rank_distribution} shows the existence of power-law scaling  (Zipf's law) in the rank distribution of visited locations in human mobility.
We also observe a convergence and robustness at the individual level, which clearly indicates non-uniform mobility behaviour and its effect on entropy, hinting at the non-Markovian nature of human mobility~\cite{zhao2015non}.\\

\noindent {\bfseries{Dwell Time Distribution.}} To further confirm this non-Markovian nature, we check the distribution of the dwell-times associated with the individual locations.  
The current mobility models are based on an assumption that human movements are randomly distributed in space and time, hence are approximated by a Poisson process~\cite{barabasi2005origin, newman2005power}.
However, Barabasi~\cite{barabasi2005origin} shows that human activities are non-Poissonian, by showing that inter-event timings depict long-tailed distribution.
We observe a similar behaviour when considering  human mobility in all the datasets, when examining the dwell-times associated with each location; most locations are visited at high periodicity, while few locations encounter long waiting times.
The current models assume that inter-event time follows exponential distribution~\cite{barabasi2005origin}, rather, we observe an emergence of power-law as shown in Figure 5, 6, 7.
The spikes in the plot correspond to delays and display the visit regularity, which indicates a long-tailed process.
The delay-time distribution depicts the priority list model in human mobility, bearing similarity to other activities as remarked by Barabasi~\cite{barabasi2005origin}.
When an individual is presented with multiple events under the context of mobility, the next location is determined on a perceived priority, thus resulting in power-law dynamics in inter-location waiting times~\cite{barabasi2005origin}.
This shows that the dwell-times associated with human mobility are not memoryless, hence cannot be considered as Markovian. 
In the above analysis, we also observe a convergence between individual mobility patterns and aggregated datasets, which concurs with the observations of Yan et al.~\cite{yan2013diversity}. 

 \begin{figure}[h!]
    \centering
      \begin{subfigure}[b]{0.58\textwidth}
        \includegraphics[scale=0.24]{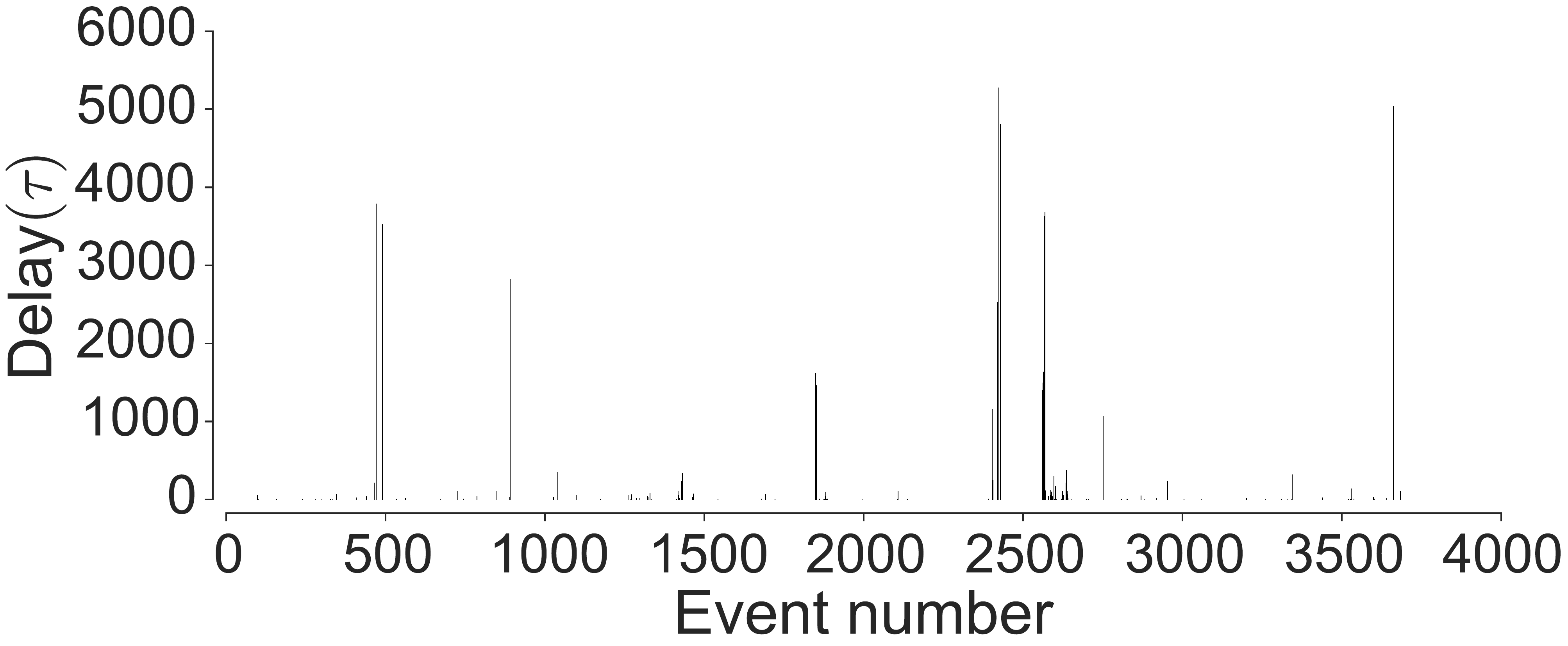} 
        \caption{Event distribution}
        \label{fig:123}
    \end{subfigure}
    \begin{subfigure}[b]{0.40\textwidth}
        \includegraphics[scale=0.16]{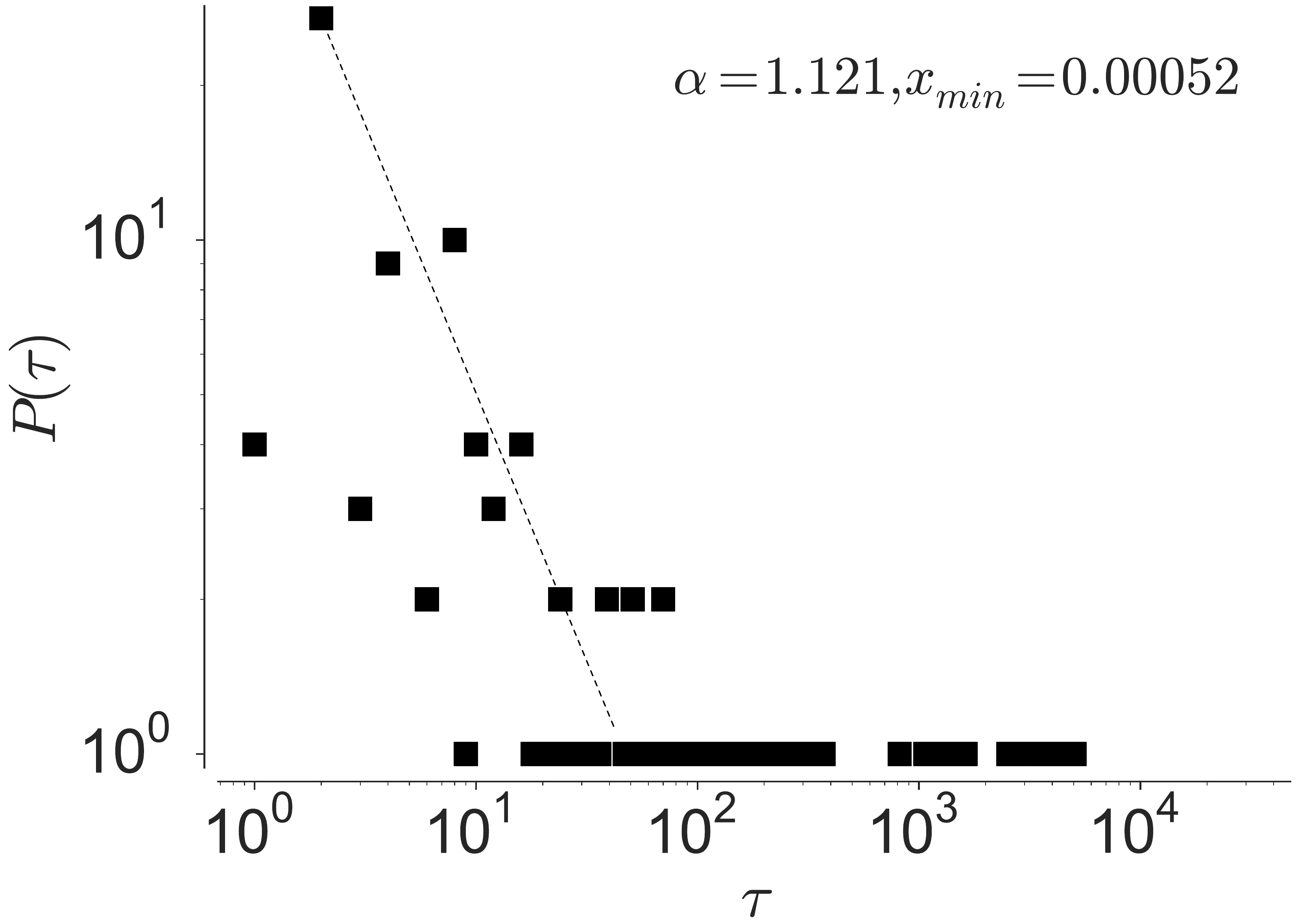} 
        \caption{Delay time distribution}
        \label{fig:fig39}
    \end{subfigure}
    \caption{PrivaMov dataset}
    \label{fig:privamov_dwelll_times}
\end{figure}

\begin{figure}[h!]
    \centering
      \begin{subfigure}[b]{0.58\textwidth}
        \includegraphics[scale=0.24]{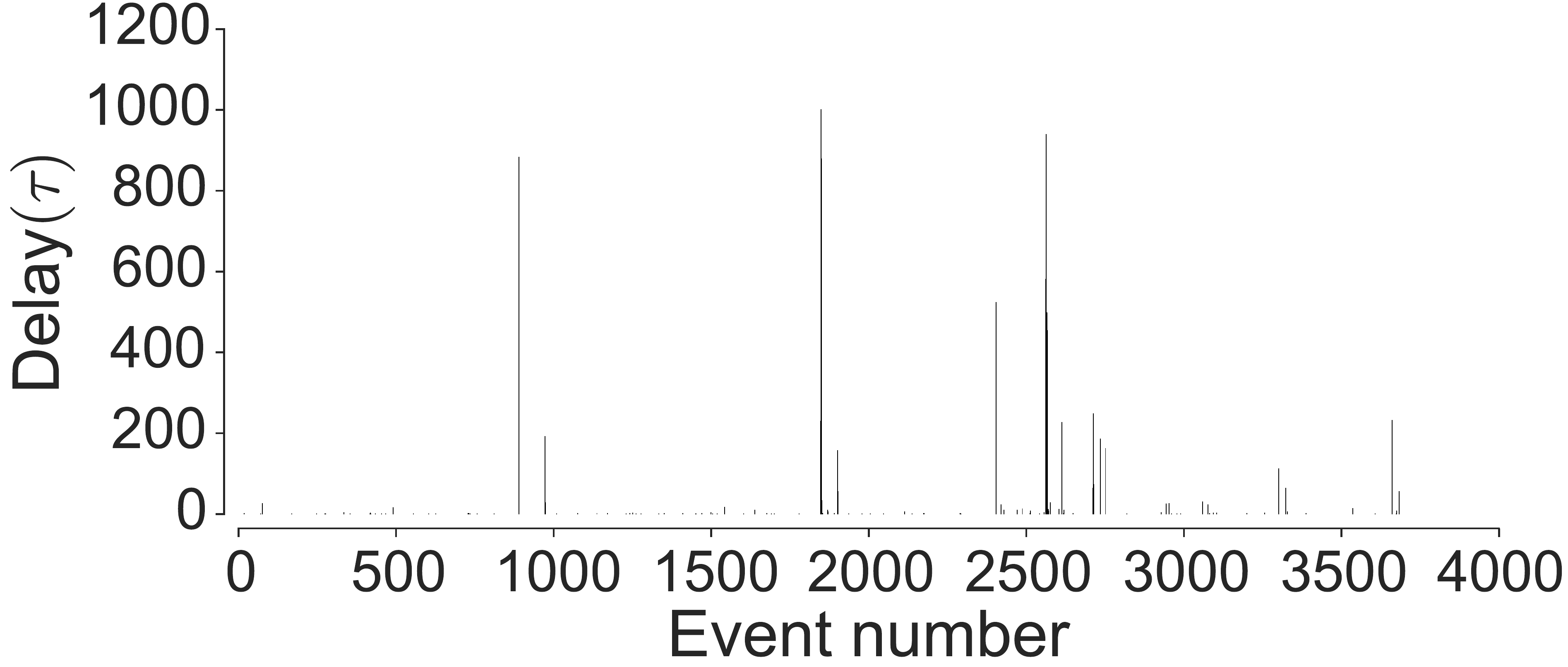} 
        \caption{Event distribution}
        \label{fig:fig31}
    \end{subfigure}
    \begin{subfigure}[b]{0.40\textwidth}
        \includegraphics[scale=0.16]{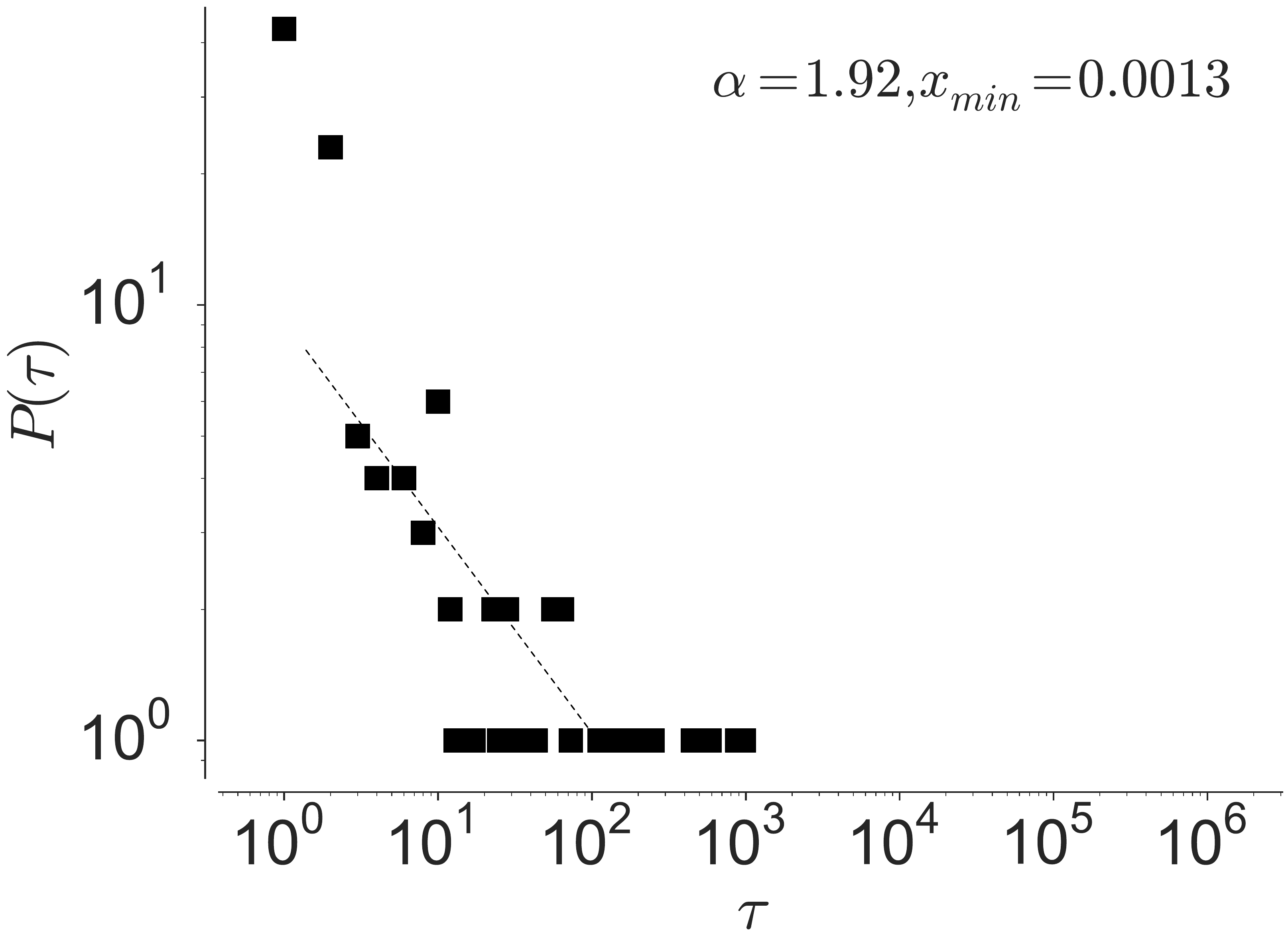} 
        \caption{Delay time distribution}
        \label{fig:fig33}
    \end{subfigure}
    \caption{NMDC dataset}
    \label{fig:nmdc_dwell_time}
\end{figure}

\begin{figure}[h!]
    \centering
      \begin{subfigure}[b]{0.58\textwidth}
        \includegraphics[scale=0.24]{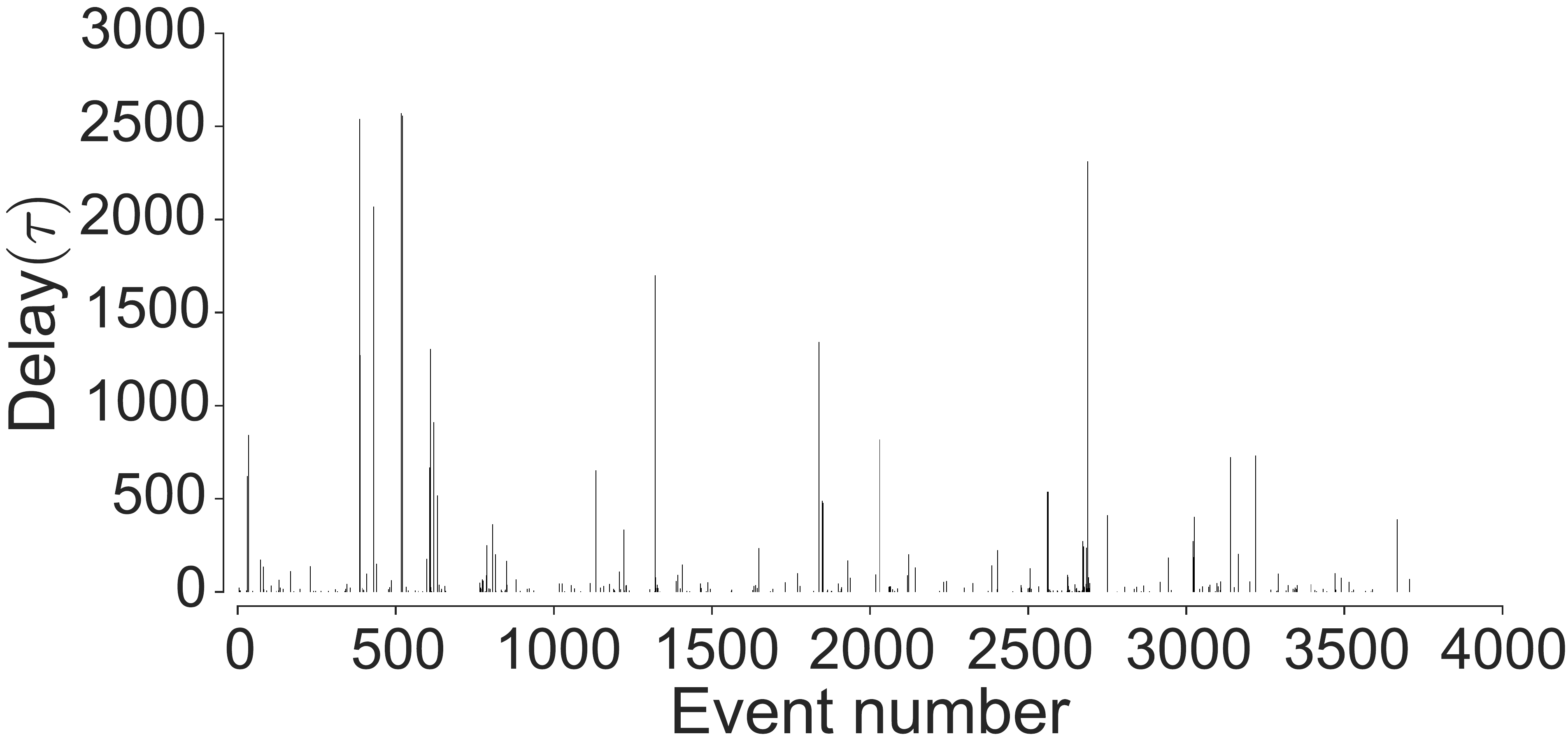} 
        \caption{Event distribution}
        \label{fig:fig32}
    \end{subfigure}
    \begin{subfigure}[b]{0.40\textwidth}
        \includegraphics[scale=0.16]{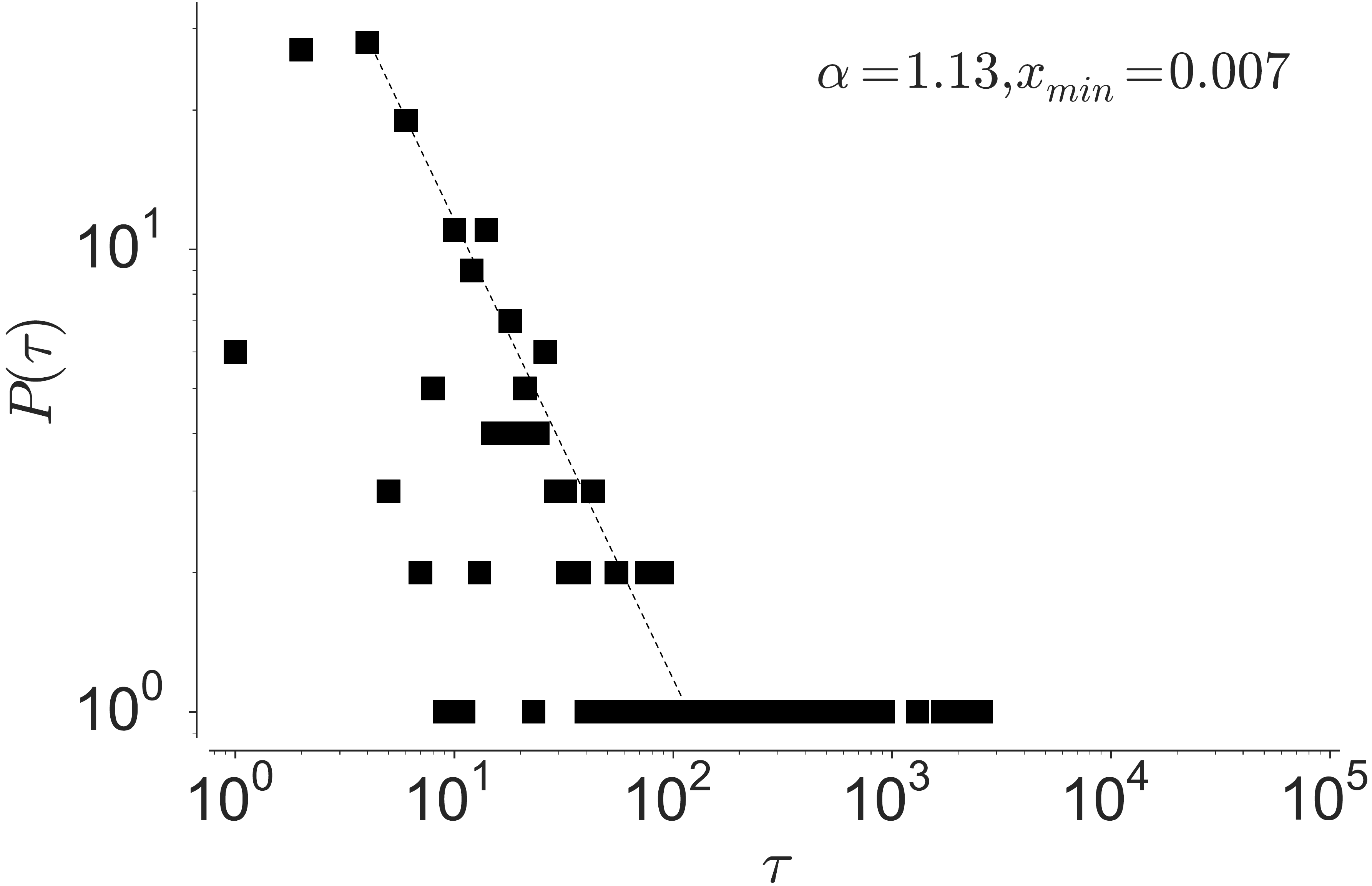} 
        \caption{Delay time distribution}
        \label{fig:fig36}
    \end{subfigure}
    \caption{GeoLife dataset}
    \label{fig:geolife_dwell_time}
\end{figure}

\noindent {\bfseries{Criticality and Mutual Information.}} Criticality is the property of dynamic systems to regulate their microscopic behavior to be spatiotemporally scale-independent~\cite{turcotte1999self}.
As a result, critical behaviour implies scale invariance; and when criticality is involved, the effects at distances much longer than microscopic lengths are crucial to study.
In practice, criticality is observed when there exists a correlation time-scale that diverges to infinity, thus the relationships in the sequence are arbitrarily non-local.

\begin{figure}[t!]
    \centering
      \begin{subfigure}[b]{0.495\textwidth}
        \includegraphics[scale=0.185]{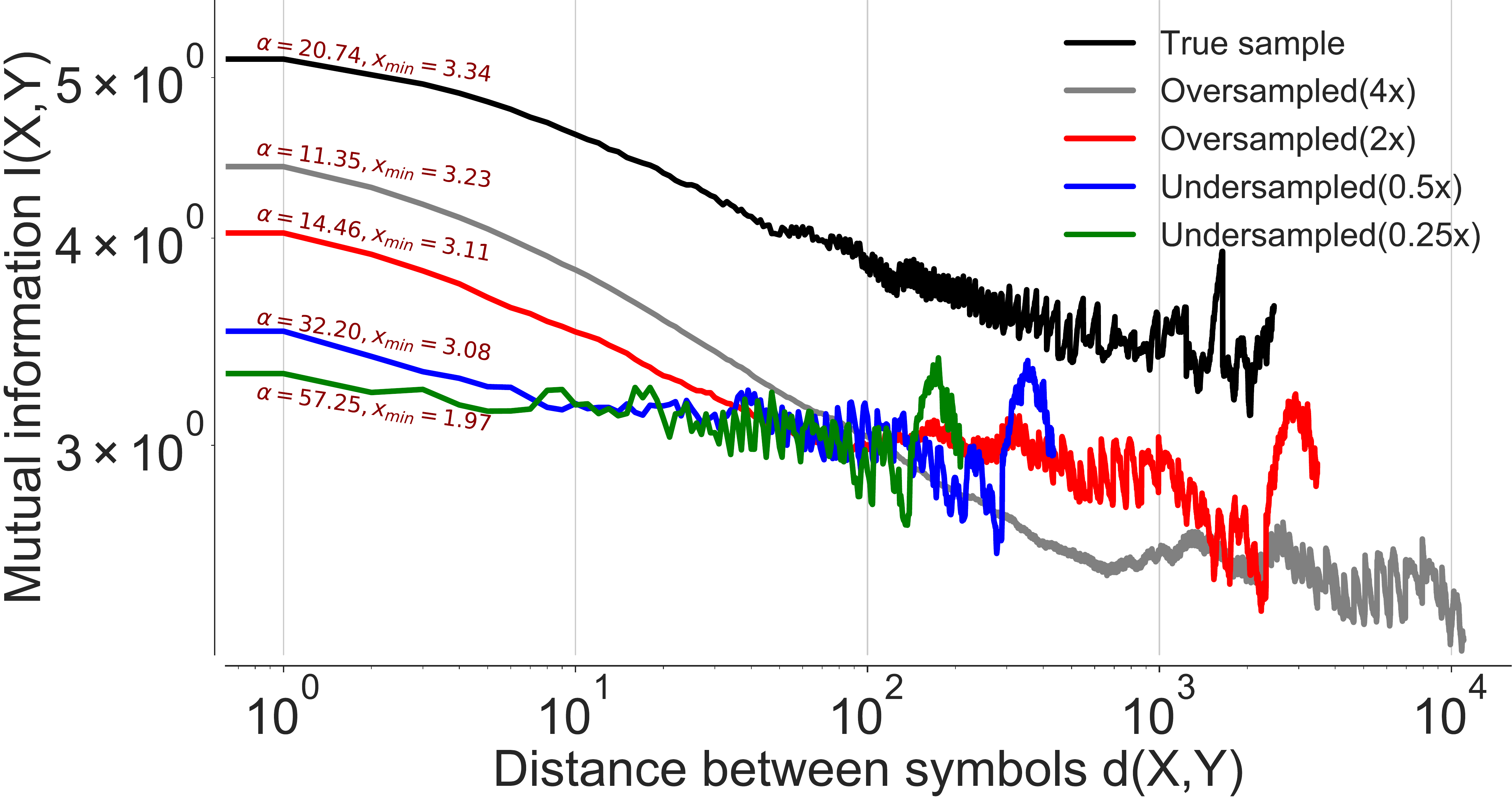}
        \caption{Mutual information decay}
        \label{fig:mutual_information_decay}
      \end{subfigure}
    \begin{subfigure}[b]{0.495\textwidth}
        \includegraphics[scale=0.17]{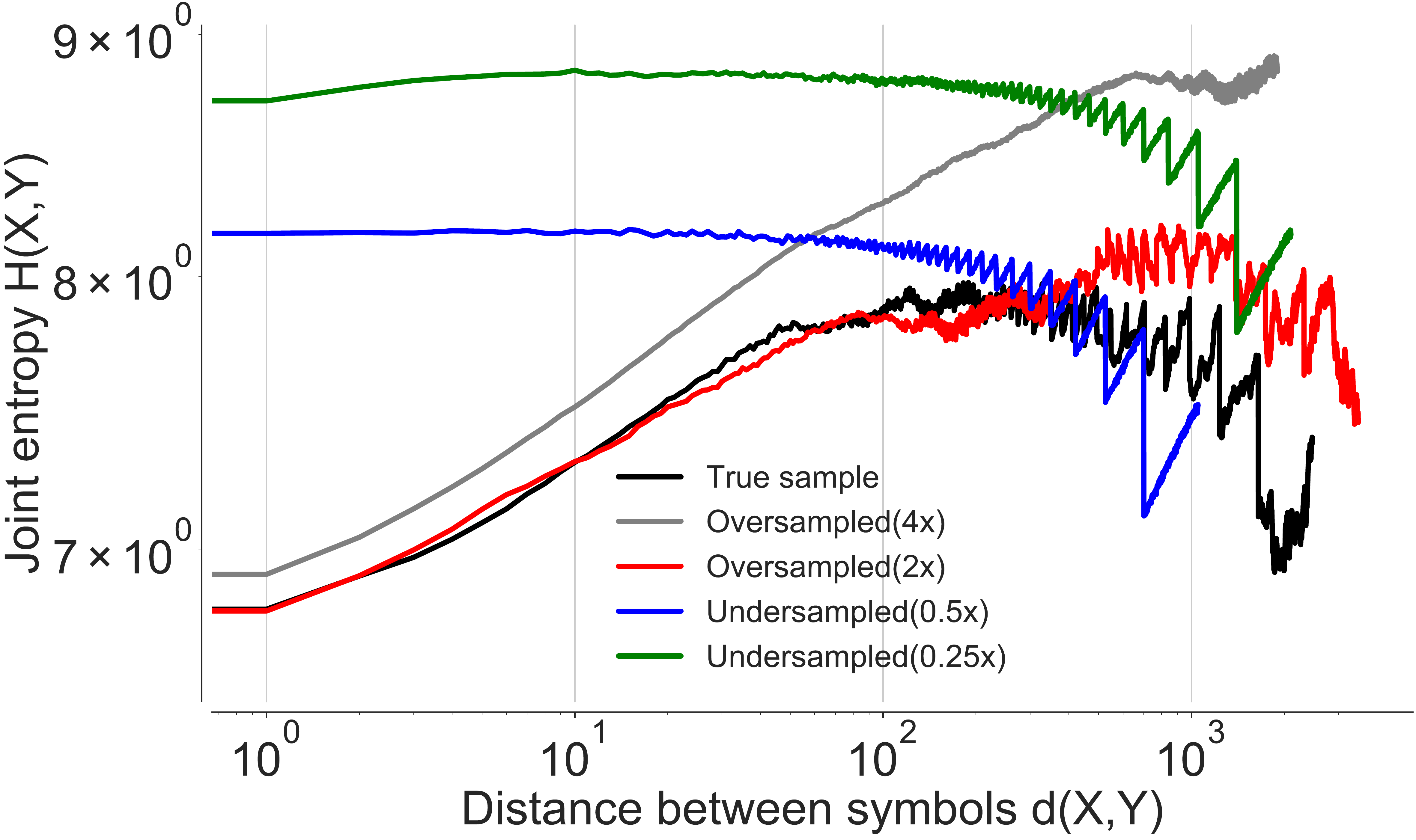}
        \caption{Joint entropy}
        \label{fig:joint_entropy}
    \end{subfigure}
    \caption{Mutual information decay for the GeoLife dataset at different sampling rates of the raw GPS coordinates projected onto a grid through Google S2~\cite{s2geo}. The upsampling is performed by the semivariance interpolation scheme~\cite{Kulkarni:2017:EHW:3139958.3140002}}.
    \label{fig:mi_geolife}
    \vspace{-10px}
\end{figure}

Mutual information $I$ quantifies the measure of information provided by a symbol/location coordinate ($Y$) about another symbol/location coordinate ($X$).
In case of mobility, mutual information between two location instances is the realisation of a discrete stochastic process, with separation $\tau$ in time~\cite{lin2016critical}.
Lin et at.~\cite{lin2016critical,cflsp} express $I$, as a function of the number of symbols (locations) between any two symbols and state that it would decay with a power-law for any context-free grammar and hence must be non-Markovian.
In order to perform this validation on human mobility, we first estimate mutual information as a function of distance for the GeoLife dataset~\cite{zheng2010geolife}.
This choice is based on uniformly sampled location points in the dataset.
To validate the emergence of power-law at distinct sampling rates, we undersample and oversampled (spatial semivariance interpolation~\cite{Kulkarni:2017:EHW:3139958.3140002}) the dataset by a factor of two and four and show the trend in Figure~\ref{fig:mutual_information_decay}.

\begin{figure}[h!]
\centering
\includegraphics[scale=0.17]{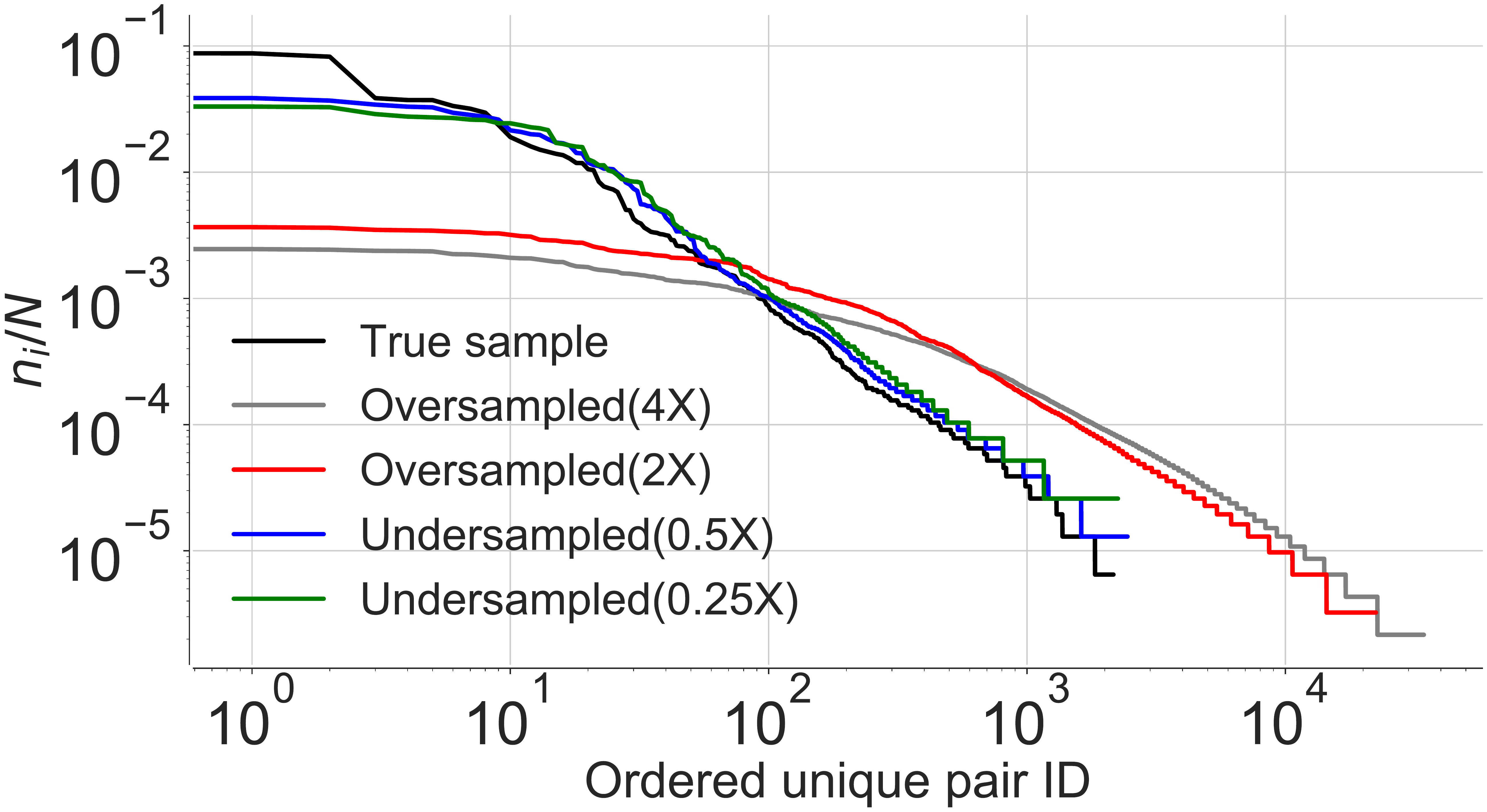}
\caption{Location pair occurrences across all the sampling rates of the true sample. The x axis represents the unique pair ID in the descending order of their frequency of occurrence. The y axis is the ratio between the unique pairs and the total number of pairs contained in the an individual trajectory.}
\label{fig:pair_occu}
\end{figure}

As suggested by Lin et al.~\cite{lin2016critical}, we observe a power-law decay at all the sampling rates.
Contrary to our assumption that $I$ would increase and decrease by the factor of under/over sampling, we observe a decrease in $I$ for all the contexts in which the true distribution of the data is altered.
We also observe that the reduction is proportional to the Kullback-Leibler divergence~\cite{PrezCruz2008KullbackLeiblerDE} between their respective distributions.
The reduction in $I$ stems from the fact that a change in the distribution results in the alteration of the true correlation between the location pairs.
The true distribution will therefore show maximum $I$, compared to the cases when either artificial pairs are introduced (oversampling) or true pairs are removed (undersampling) from the dataset.
To verify our hypothesis, we also calculate the joint entropy for all the cases and observe an increase in $H(X,Y)$ for the altered distributions as shown in Figure~\ref{fig:joint_entropy}. 
We see that the increased entropy is due to an increase in the ratio between unique pairs in the dataset over the total number of pairs.
The introduction of spurious pairs scrambles the true distribution as it leads to introduction of data points in the true sequence, thereby changing the random variables sampled at distance $D$, hence reducing $I$.
This occurrence was confirmed after computing the area under the curve (ROC), which was maximum for the true data distribution in the first quartile as compared to the rest as shown in Figure~\ref{fig:pair_occu}. 
This explains our observation of higher joint entropy for the oversampled and the undersampled case.

\begin{figure}[h!]
    \centering
      \begin{subfigure}[b]{0.54\textwidth}
        \includegraphics[scale=0.20]{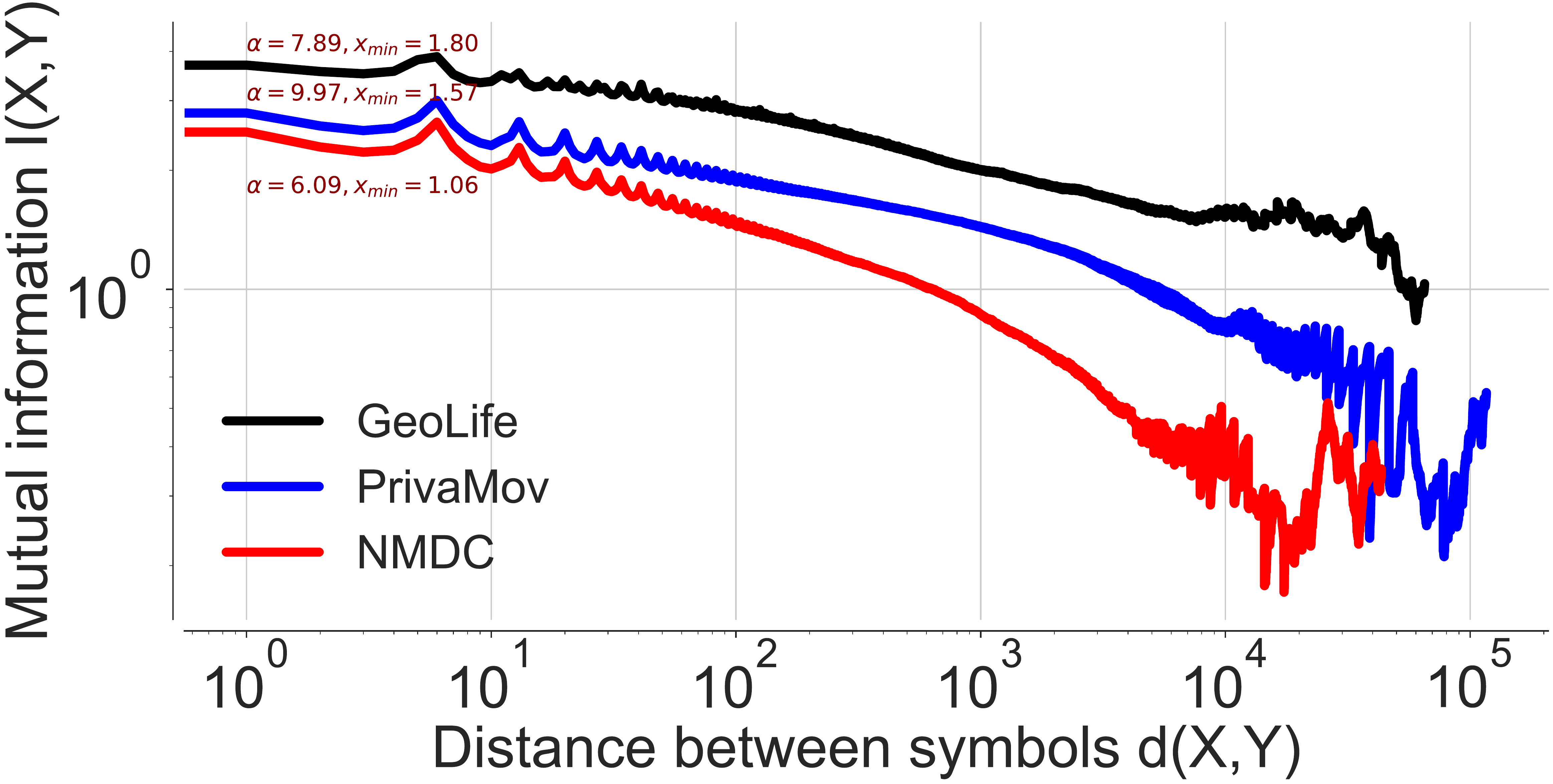}
        \caption{Mutual information decay}
        \label{fig:fig41}
    \end{subfigure}
    \begin{subfigure}[b]{0.44\textwidth}
        \includegraphics[scale=0.18]{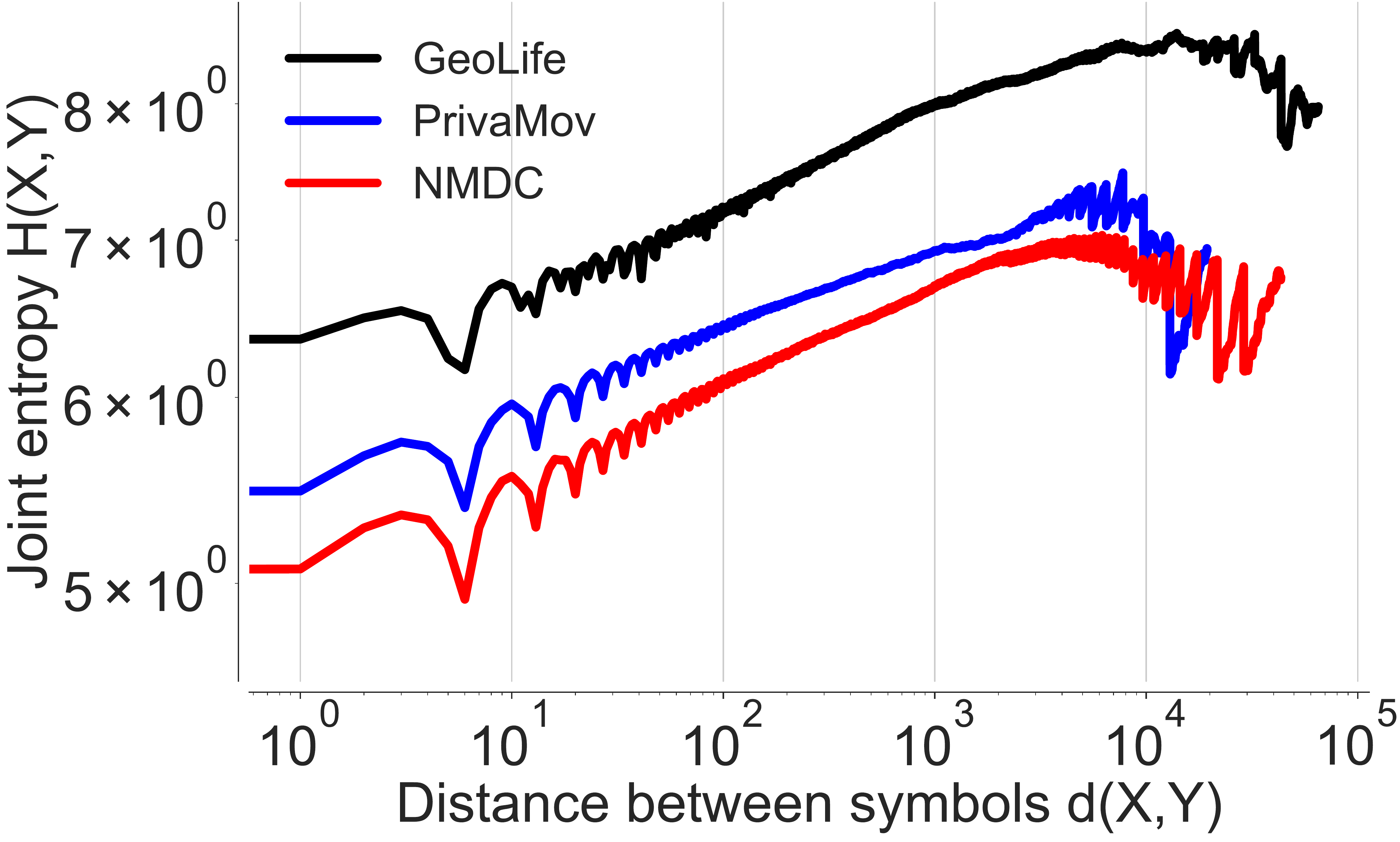}
        \caption{Joint entropy}
        \label{fig:42}
    \end{subfigure}
    \caption{Mutual information decay and joint entropy estimated for all the datasets. The dataset consists of stacked sequences of temporally arranged individual points of interest.}
    \label{fig:mi_decay_all}
\end{figure}

To analyse the long-range correlations present in each of the datasets, we compute their respective mutual information decay.
This information will serve as basis for the difference in accuracy for each dataset and the performance difference between the prediction algorithms.
Again, we observe a power-law decay across all the datasets and their respective joint entropy, as shown in Figure~\ref{fig:fig41} and Figure~\ref{fig:42}.
We further explore the Markov transition matrices for these datasets and observe that they are reducible and periodic, resulting in the decay of $I$ to a constant.
It has been shown that such a characteristic of the transition matrix cannot result in an exponential decay by Lin et al.~\cite{lin2016critical, cflsp}.  
They show that an irreducible and aperiodic Markov process, with non-degenerate eigenvalues, cannot produce critical behaviour because $I$ decays exponentially. 
This phenomenon is seen in a number of cases, including hidden and semi-Markov models~\cite{lin2016critical, cflsp}.
In the literature, such behaviour is superficially dealt with by increasing the state space to include symbols from the past, which does not address the main issue~\cite{cflsp} with Markov models; lack of memory.
This analysis shows that GeoLife dataset consists of considerably higher number of long-range correlations, compared to the PrivaMov dataset and the NMDC dataset.
This should be self-evident from their respective data collection durations. 
However, the lower dependencies in the NMDC dataset, compared to PrivaMov, is due to the smaller area of the data collection region, which generally results in lower entropy of movement~\cite{lu2013approaching, Song2010LimitsOP}.

Here, we reason about the accuracy variation within and between the datasets and about the performance differences between the prediction algorithms. 
We observe that the NMDC dataset provides higher accuracy as compared to the other datasets, and witness a lower variation within the accuracies of different algorithms. 
This stems from the presence of very short dependencies in the individual trajectories present in the dataset, as seen in Figure~\ref{fig:fig41}.
The lower correlations also result in roughly equivalent prediction accuracies within the predictive models. 
The lower accuracies of recurrent-neural architectures, compared to Markov chain at some time-steps are due to the models tendency to actively seek for long-range dependencies.
However, if the dataset does not contain such dependencies, the model underperforms, unless it is weighted to account for such an existence. 
This underperformance is evident from the behaviour of dilated-RNN's, where an increase in dilations (to account for longer dependencies) results in dropping accuracy.
Such a phenomenon has also been observed in language modelling tasks, which suggests that this is not a domain specific occurrence~\cite{khandelwal2018sharp}.
The performance drop in the recurrent-neural architectures at different time steps is due to capturing the long-distance dependencies to different degrees, resulting in either under/over fitting.
An additional reason for higher accuracy in NMDC dataset is due to a lower number of unique locations and smaller variations in the dwell-times, as compared to the PrivaMov and GeoLife datasets, as shown in Figure~\ref{fig:rank_distribution} and Figure~\ref{fig:nmdc_dwell_time}.
These aspects directly correlate with the entropy and thus affect predictability~\cite{Song2010LimitsOP}.
We also observe that PSMMs, perform better on GeoLife dataset, compared to other two, due to its ability to search for dependencies at longer distances.\\

\noindent {\bf{Entropy and Predictability Estimation.}} The current method~\cite{Song2010LimitsOP, lu2013approaching, ikanovic2017alternative, zhao2015non} uses Lempel-Ziv data compression scheme~\cite{ziv1978compression} to compute the mobility entropy. 
This approach segments the complete trajectory into substrings, where a substring is defined as the shortest length element subsequence yet to be encountered.
As observed by Lesne et al.~\cite{lesne2009entropy}, a vast majority of substrings are of length one or two, which are the dominant contributors to the entropy.
The estimated entropy is thus the outcome of finite-size fluctuations; and the total count of the substrings and of the elements in a substring does not represent the true probability distribution. 
Furthermore, in this process the structural correlation between the individual substrings is ignored, based on the argument that the probability of joint occurrences is weak~\cite{lesne2009entropy}.
This argument stems from the reasoning that the parsed substrings are independently and identically distributed (iid) according to Gaussian distribution, that does not apply to mobility trajectories.
Finally, the correlated features can be compressed only by memorising all the cases of intervening random variables between the correlated instances.~\cite{storer1987data}.
It has thus been proved that Lempel-Ziv approach fails to capture redundancies in the data sources with long-range correlations~\cite{lesne2009entropy}.

More importantly, Storer et al.~\cite{storer1987data} shows that standard data compression approaches, such as the Lempel-Ziv approach cannot truly capture long-range dependencies, as the information carriers of a sequence lie in its structural origin. 
However, these approaches limit the entropy estimation process at the sub-string level.
Given that entropy is the complete quantitative measure of the dependency relations (including many point correlations), the computation of higher-order entropy is non-trivial. 
Therefore, it is flawed to assume that the $\pi^{max}$ derived from such an approximate estimation of $S^{real}$ should act as an upper bound of predictability on trajectories compiled for long time-spans.

\section*{Discussion}

The previous research~\cite{Song2010LimitsOP, lu2013approaching} estimated $S^{real}$ and $pi^{max}$ by using CDR datasets spanning a period of three to five months.
Such datasets do not truly capture features such as the total number of unique locations visited by an individual, due to its low granularity (typically 4-5km~\cite{ikanovic2017alternative}).
This results in a dataset with a masked entropy and mobility patterns ignoring long-range correlations. 
An important point to note is that for very short distances, power-law decay and exponential decay may not be trivial to differentiate~\cite{newman2005power}.
This was due in part to the unavailability of high granularity GPS datasets and the fact that that previous works~\cite{Song2010LimitsOP, lu2013approaching} were only studied for short distances of human mobility. 
Therefore, the assumptions underlying the computation of $S^{real}$ and $\pi^{max}$ would have been fairly easy to overlook.

The aforementioned inadequacies would reinforce the empirical validation of $\pi^{max}$ using Markov chains, however, as mentioned above this would result in an error-prone estimation of the predictability. 
As seen in other domains of sequential-data modelling such as natural language processing, it has been shown that Markov chains are fundamentally unsuitable for modelling such processes~\cite{chomsky1959certain}.
Our empirical observations, backed by theoretical foundations, indicate that human mobility will be poorly approximated by Markov chains.
This is particularly afflictive for trajectories that satisfy criteria such as long time-span of collection and large radius of gyration of movement.

Our choice to rely on mutual information was based on its triviality and its domain independence.
As shown by Lin et al.~\cite{lin2016critical}, the mutual information decay offers some insights into why recurrent-neural architectures exceed probabilistic models in terms of capturing criticality. 
As for a Markov process, the observations at $t_n$ depends only on events at previous time step $t_{n-1}$ or on previous $n$ time-steps for an $n$-order Markov chain.
Under such a context, the maximum possible predictive information is given by the entropy of the distribution of states at one time step, which is in turn bounded by the logarithm of the number of accessible states.
Unlike Markov chains, the recurrent-neural architectures, such as RHN's, approach this bound while maintaining the memory long enough, that the predictive information is reduced by the entropy of transition probabilities.

We expect to provide a more sophisticated description of the underlying phenomenon as more of the trajectory is observed. 
Consequently, increasing the number of parameters in the model. 
That is, when we examine trajectories on the scale of individual coordinates, we learn about the rules of combining these points into points of interest and the transition paths between them.
At the next level, if we consider several of these points of interest and the paths, we learn the rules for combining these points into semantic patterns.
Similarly, when we look at semantic patterns, we learn about the visitation periodicities and circadian rhythms associated with the mobility behaviours.
Therefore, longer traces have increasing number of long range structural correlations that are non-trivial to be captured by the currently available entropy measure.
Moreover, the current approximation implies that the substrings have the same compressibility factor~\cite{ziv1978compression}, hence the results derived from this approach would coincide with the average.   
Thus, the current computation will result in higher estimates of entropy, consequently resulting in a lower predictability bound.

Even though Markov models tend to underperform in modelling human mobility, their use in human mobility prediction is not entirely without precedent.   
In fact, considering their low computational complexity, it might be advantageous to opt for a Markov model when a dataset contains short-distance dependencies and low number of unique locations.
However, in datasets exhibiting criticality, long-range correlations appear in the vicinity of the critical point, which necessitate recurrent-neural architectures to accurately model human mobility. 
In this paper, we have shown that human mobility exhibits scale-invariant long-range correlations which can be quantitatively measured by a power-law decay of mutual information.
We highlight that the exponent characterising the power-law decay of the correlations is well defined for infinite sequences. 
For mobility trajectories, however, the accuracy of the analysis is restricted by the length of the substrings and their entropy, which results in an incorrect estimation of maximum predictability.
This explains why the empirical results surpass the theoretical upper bound in several previous research works and in our own experiments.

\section*{Methods}

{\bfseries{Mobility prediction.}} We define mobility prediction as forecasting the transitions between places, after eliminating all self-transitions~\cite{cuttone2018understanding, smith2014refined}.
A preliminary step in achieving this consists of transforming the raw GPS locations into a sequences of points of interest~\cite{Kulkarni:2017:EHW:3139958.3140002}.
A point of interest is defined as any location where an individual visits with an intentional purpose with a perceived priority for e.g., home/work place, gym, train station etc. 
Among plethora of existing works dedicated to the problem of extracting these points, we rely on our approach that is independent of {\it{a priori}} assumptions regarding the data and individual mobility behaviors~\cite{Kulkarni:2017:EHW:3139958.3140002}.
We then convert the raw GPS trajectory of a user $u$, $T_u = \langle(lat_1,lon_1,t_1), (lat_2,lon_2,t_2)...(lat_n,lon_n,t_n) \rangle$, where $lat_i, lon_i$ are the latitude and longitude coordinates respectively and $t_i$ is the timestamp such that $t_{i+1} > t_i$ into a sequence of temporally ordered points of interest, $s(t) = \langle (poi_1,t_1), (poi_2 t_2)...(poi_n,t_n)\rangle$, where $poi_i$ is the point of interest at index $i$.
The mobility prediction task is thus formulated as: given a sequence $s(t)$ up to a timestamp $n$, predict the next point of interest at timestamp $n+1$. 
The prediction accuracy is then estimated by following the approach stated by Lu et al.~\cite{lu2013approaching}.\\

\noindent {\bfseries{Predictive models.}} We use the standard implementations of the predictive algorithms as described in their respective papers. 
Markov chains~\cite{gambs2012next} and hidden Markov models~\cite{si2010mobility} are implemented using the standard python libraries (hmmlearn).
We use hyper-parameters stated in these papers.
Vanilla-RNN~\cite{grossberg2013recurrent}, RNN-LSTM~\cite{Hochreiter1997LongSM} and dilated-RNN~\cite{Chang2017DilatedRN} are based on predicting the next character (language modeling) in the text, whereas RHN~\cite{Zilly2017RecurrentHN} and PSMM~\cite{Merity2016PointerSM} model the prediction task as multivariate classification. 
For dilated-RNN~\cite{Chang2017DilatedRN} we use the dilations of 1, 2, 4, 8, 16, 32 and 64 and provided the results for dilation 32 after which we observe a drop in the accuracy.\\

\noindent {\bfseries{Computing $S^{real}$ and $\pi^{max}$.}} These values are computed as stated in Song et al.~\cite{Song2010LimitsOP} as per Equation~\ref{eq:eq1} which is based on Lempel-Ziv data compression~\cite{ziv1978compression}.

\begin{equation}
S^{real} = (\frac{1}{n}\sum_{i=1}^{n}\lambda_i )^{-1}ln(n)
\label{eq:eq1}
\end{equation}

Where, $n$ is the length of the trajectory (total number of locations) and $\lambda$ is defined as the length of the shortest substring at an index $i$ not appearing previously from index 1 to $i-1$.
Note that we use the same base (2) in entropy estimation as for the logarithm in Fano's inequality.
Furthermore, their length is set to zero upon reaching index $i$, when no more unique substrings can be computed using the above method. 
$\pi^{max}$ is then estimated by solving the limiting case of Fano's inequality~\cite{gerchinovitz2017fano}.\\

\noindent {\bfseries{Mutual information.}} Equation~\ref{eq:mi} gives mutual information, $I$, between two discrete random variables $X,Y$ jointly distributed according to the probability mass function $p(x,y)$.
Mutual information $I$ thus quantifies the number of bits of information provided by a symbol ($Y$) about another symbol ($X$).

\begin{equation}
\label{eq:mi}
I(X,Y) = \sum_{x,y}p(x,y)\log(\frac{p(x,y)}{p(x).p(y)})
\end{equation} 

Mutual Information can also be estimated by computing the entropy of the marginal distribution of discrete random variables $X$ and $Y$, and the joint entropy of discrete random variables $X$ and $Y$.

\begin{equation}
I(X,Y) = H(X) + H(Y) - H(X,Y) = D_{KL}(p(XY)||p(X)p(Y))
\end{equation}

where \emph{H}(\emph{X}) is the entropy of a random variable \emph{X} and \emph{H}(\emph{X, Y}) is the joint entropy of \emph{X} and \emph{Y}. 
$D_{KL}$ is the Kullback-Liebler Divergence~\cite{PrezCruz2008KullbackLeiblerDE}. 
Thus, Mutual Information is same as the Kullback-Leibler Divergence between distributions of \emph{X} and \emph{Y}.
In order to compensate for insufficient samplings, we use the following adjustment proposed by Grassberger et al.~\cite{grassberger2003entropy} to compute \emph{H}(\emph{X}), \emph{H}(\emph{Y}), \emph{H}(\emph{X,Y}).
\begin{equation}
H(X) = \log N - 1/N \sum_{i=1}^{k} N\textsubscript{i} \psi((N\textsubscript{i})
\end{equation}

In order to compute the mutual information in a mobility trajectory, we first estimate the distribution of a random variable from index 0 followed by the distribution of the second random variable in the data at index $D$. 
The random variables \emph{X} and \emph{Y} are sampled from the individual trajectory sequence. 
\emph{X} denotes sequence starting at index 0 and \emph{Y} is sequences starting at offset \emph{D} from 0. 
\emph{D} is then varied to compute long-distance dependencies at every separation by creating displacements between the random variables. 
Once the contextual dependence limit is reached, this approach starts sampling noise, which sets the termination criterion.
Finally, the average similarity between the two symbols is quantified.

\section*{Data availability}

The GeoLife dataset that supports the findings of this study is public and is made available by Microsoft Asia~\cite{geolife}. 
The PrivaMov dataset is collected by Universite de Lyon and can be obtained by submitting an online form~\cite{privamov}. 
The Nokia Mobile Data Challenge (NMDC) can be obtained in a similar fashion~\cite{nmdc}.
We provide links in the citations to avail these datasets.
Furthermore, our source codes are made public~\cite{ldd} and further clarifications will be provided upon request.
\bibliographystyle{abbrv}
\bibliography{biblio}

\end{document}